\documentclass[twocolumn]{aastex7}

\begin{document}

\title{The Accretion Disk Size Problem in AGN Disk Reverberation Mapping is an Obscuration Effect: A Uniform AGN Sample Study with Swift}

\author[0000-0002-8671-1190]{Collin Lewin}
\affiliation{MIT Kavli Institute for Astrophysics and Space Research, MIT, 77 Massachusetts Avenue, Cambridge, MA 02139, USA}
\email{clewin@mit.edu}

\author[0000-0003-0172-0854]{Erin Kara}
\affiliation{MIT Kavli Institute for Astrophysics and Space Research, MIT, 77 Massachusetts Avenue, Cambridge, MA 02139, USA}
\email{}

\author[0009-0001-9034-6261]{Christos Panagiotou}
\affiliation{MIT Kavli Institute for Astrophysics and Space Research, MIT, 77 Massachusetts Avenue, Cambridge, MA 02139, USA}
\email{}

\author[0000-0002-8294-9281]{Edward M.\ Cackett}
\affiliation{Department of Physics and Astronomy, Wayne State University, 666 W.\ Hancock St, Detroit, MI, 48201, USA}
\email{}

\author[0000-0001-9092-8619]{Jonathan Gelbord}
\affiliation{Spectral Sciences Inc., 30 Fourth Ave. Suite 2, Burlington MA 01803}
\email{}

\author[0000-0002-6733-5556]{Juan V.\ Hern\'{a}ndez Santisteban}
\affiliation{SUPA School of Physics and Astronomy, North Haugh, St.~Andrews, KY16~9SS, Scotland, UK}
\email{}

\author[0000-0003-1728-0304]{Keith Horne}
\affiliation{SUPA School of Physics and Astronomy, North Haugh, St.~Andrews, KY16~9SS, Scotland, UK}
\email{}

\author[0000-0002-2180-8266]{Gerard A.\ Kriss}
\affiliation{Space Telescope Science Institute, 3700 San Martin Drive, Baltimore, MD 21218, USA}
\email{}

\begin{abstract}
In the past decade, Swift has performed several AGN high-cadence reverberation mapping campaigns, and generally found that the UV/optical interband lags  are $\sim$3 times longer than predicted for a standard thin disk, thus coined ``the accretion disk size problem". Here we present a systematic sample of Swift-monitored AGN. In this analysis, we confirm the accretion disk size problem, but find that the lag excess occurs only in the subset of obscured AGN, which show a significantly elevated mean normalization of $5.21 \pm 0.47$ ($p = 0.008$), whereas the unobscured AGN exhibit a mean excess consistent with standard disk predictions ($1.00 \pm 0.31$). Correlation and regression analyses similarly reveal X-ray column density as the strongest predictor of lag excess, explaining over 80\% of its variance. We interpret these results as line-of-sight obscuration being linked to the too-long lags via additional reprocessed emission from the absorbing material itself. The consistency of lags in the unobscured subgroup with standard disk predictions suggests that the accretion disk size problem is not the result of shortcomings of standard accretion disk theory nor contamination by the broad-line region (BLR). X-ray to UV lag amplitudes and correlations show more complex and variable behavior in obscured AGN, suggesting that obscuration may disrupt or complicate the connection between high- and low-energy emission potentially through reprocessing, scattering, and/or ionization changes.
\end{abstract}

\section{Introduction} \label{sec:intro}

Active galactic nuclei (AGN) are powered by accretion onto a supermassive black hole, a process that releases an enormous amount of energy that drives feedback on a scale that affects galactic evolution \citep[e.g.,][]{Fabian_2012}. The innermost regions of AGN--the accretion disk and broad-line region, or BLR--are located at sub-parsec scales and are thus unresolvable in almost all cases, except a few nearby AGN have had their inner structure partially resolved via infrared interferometry \citep[e.g.,][]{Gravity_2018, Gravity_2020}. For the vast majority, we must instead infer the structure through spectral and timing information, the latter mainly focused on a technique known as reverberation mapping \citep{Blandford_1982, Peterson_2004}.

Reverberation mapping capitalizes on estimating the light-travel time between the central engine and surrounding emitting regions: variations in the X-ray emission from close to the black hole by the corona are ``echoed" after a time delay as the emission is reprocessed by farther regions. By measuring the time lags between the correlated variability that arises in different wavebands, one can thus constrain the relative distances of the reprocessing regions from the black hole. This technique has been widely and successfully applied to map the BLR via emission line responses to continuum changes \citep[][and references therein]{Peterson_2004, Bentz_2009} to measure black hole masses. More recently, reverberation mapping has been extended to probing the temperature-radius profile of the accretion disk out to the inner BLR and dusty torus in the the ultraviolet, optical, and infrared (UVOIR) \citep{Cackett_2021}.

In the canonical picture, when X-ray variations from the corona irradiate the disk, the disk heats up and re-emits at longer UVOIR wavelengths. Longer wavelengths originate from cooler radii located farther out, and so the lag increases with wavelength. Under the assumption of a standard disk with a temperature profile $T(r)\propto r^{-3/4}$, the expected lag as a function of wavelength follows a power-law $\tau(\lambda)\propto \lambda^{4/3}$ \citep{Collier_1998, Collier_1999, Cackett_2007}. The normalization of this lag–wavelength relation is set by the light-crossing time of the disk’s thermal emission region, which depends on the black hole mass, accretion rate, and disk structure and properties \citep[e.g., opacity, geometry;][]{Fausnaugh_2016}. As a result, measuring the disk reverberation lags and comparing to this baseline model provides a test of accretion disk theory.

Over the past decade, intensive multi-wavelength monitoring campaigns have greatly expanded our catalog of UV/optical lag measurements in AGN. High-cadence, months-long programs using the \textit{Neil Gehrels Swift Observatory} and complementary ground-based optical telescopes have been carried out for on the order of ten distinct AGN \citep[e.g.,][]{McHardy_2014, Shappee_2014, Edelson_2015, Fausnaugh_2016, Cackett_2018, McHardy_2018, Edelson_2019, Cackett_2020, Hernandez_2020, Kara_2023}. These reverberation mapping campaigns of well-studied AGN broadly confirm the qualitative expectation of lags increasing with wavelength; however, a striking outcome has been that the observed lags are systematically longer than predicted by standard accretion theory, typically requiring a normalization larger by a factor of about 2–3 \citep[e.g.,][]{Edelson_2019, Cackett_2020, Hernandez_2020}. In Mrk~335, the lags were a full order of magnitude longer than expected \citep{Kara_2023}. The lags in the \textit{U} band near 3500~\AA\ are especially long, often exceeding even the best-fit lag-wavelength relation by over a factor of 2 \citep[see Figure~5 in][]{Edelson_2019}.  If the lags are attributed solely to the accretion disk, then these measurements imply that the disk is significantly larger than predicted. Such “too-long” lags indicate that our standard disk reprocessing picture is incomplete, and we henceforth refer to this discrepancy as the ``accretion disk size problem."

The UV to X-ray lags and correlations also give rise to additional open questions. The UV and X-ray light curves are also often weakly correlated or completely uncorrelated \citep[and notably less so than the UV and optical; e.g.,][]{Schimoia_2015, Starkey_2017, Buisson_2018, Cackett_2020, Cackett_2023}. The X-ray variability is sometimes seen to lag the UV \citep[e.g. in Mrk 335;][]{Lewin_2023}. These findings challenge the standard reprocessing picture, where X-ray variations are expected to drive longer-wavelength variability. One possibility is that the CCF fails to account for a dynamic corona that lowers the measured X-ray/UV correlation \citep{Panagiotou_2022a}.

A leading solution to the accretion disk size problem is the presence of additional reprocessing from the BLR, which responds on longer timescales than the inner disk, thus elongating the measured lags as the effective radius of reprocessing is larger than it would be from the disk alone \citep{Korista_2001}. Notable evidence for this theory was from Hubble Space Telescope spectroscopy of NGC~4593, which revealed a well-resolved, distinct jump in the lag spectrum at the Balmer break (3650~\AA), as expected from a separate emission component from the diffuse continuum of the BLR \citep[or any line emitting gas, like a wind][]{Cackett_2018, Lawther_2018, Korista_2019, Netzer_2022}.

The problem has been further studied using frequency-resolved analyses that offer a  complementary approach to traditional time-domain methods by explicitly isolating the variability (and thus lag) as a function of timescale. This allows us to isolate the variability on different timescales, which can result from distinct physical processes. In recent applications to NGC~5548, Mrk~335, and Mrk~817, isolating the high-frequency (i.e., short-timescale) variability revealed lags consistent with those expected from standard accretion disk theory \citep{Cackett_2022, Lewin_2023, Lewin_2024}. The lags agreed with the disk when isolating frequencies corresponding to timescales shorter than the light-crossing time of the BLR, as inferred from the H$\beta$ lag, effectively filtering out the BLR contribution.

Several other studies have sought to isolate short-timescale variability by removing long-term trends from the light curves through detrending methods, such as subtracting low-order polynomials or moving averages \citep[e.g.,][]{McHardy_2018, Hernandez_2020, Pahari_2020, Vincentelli_2021}. In doing so, these approaches have similarly recovered lag amplitudes more consistent with standard disk expectations. However, while effective, such methods do not explicitly define the timescales being filtered out, making it difficult to robustly interpret the physical origin of the remaining signal. While this evidence has been viewed as the observed reverberation being a composite of a fast disk-reprocessing and a slower reverberation from larger radii, others argue that the lag excesses are actually due to shortcomings of how the standard thin-disk model is implemented to model the lags; for instance, disk geometry \citep[e.g.,][]{Starkey_2023}, or a lack of general relativity, black hole spin, corona height, and so on in simple analytical comparisons \citep{Kammoun_2021a}.

Unique clues to solving these puzzles have come from recent dramatic, transient obscuration events that affect the measured reverberation signal. The AGN~STORM~2 campaign on Mrk~817 revealed several of these events: the source experienced heavy X-ray obscuration by a multi-phase wind, suppressing its X-ray flux \citep{Kara_2021, Partington_2023, Zaidouni_2024}. Despite this, the UV and optical continua still varied and exhibited measurable lags \citep{Cackett_2023}. Unlike anything we had seen before, the lags changed with the column density of the disk wind \textit{twice}. Epochs with heavy obscuration revealed lags that were twice as long--equivalent to a $>10\sigma$ change in the measured lag normalization. In contrast, epochs with low X-ray column density were consistent with the standard disk predictions, including a resolution of the U-band excess \citep{Lewin_2024}. Similar behavior was seen in the \ion{C}{4} lag, which was longest when the UVW2/HST correlation weakened and the column density was relatively high \citep{Homayouni_2023_2}. These findings highlight the potential role of obscuration in the continuum lag puzzle. 

In this work, we present a systematic study of the UVOIR reverberation lags across a sample of nine AGN from previous, multi-band monitoring campaigns. We investigate whether the degree that the observed lags exceed standard disk predictions correlate with physical properties of the AGN, including line-of-sight obscuration inferred by fitting the X-ray spectra.

\section{Sample and Observations}
\begin{deluxetable*}{lcccccr} \label{tab:source_properties}
\tablecaption{Properties of the AGN sources and their observations.}
\tablehead{
\colhead{Source} & \colhead{$\log_{10}(M_{BH}/M_\odot)$} & \colhead{$\dot{m}$} & \colhead{\shortstack{Date Range (MJD)}} & \colhead{\shortstack{Duration (days)}} &
\colhead{\shortstack{$\overline{\Delta t}$ (days)}} &
\colhead{Reference}
}
\startdata
Fairall 9 (Early) & $8.41^{+0.08}_{-0.12}$ & 0.035 & 58251.7--58529.2 & 277.49 & 1.24 & HS20 \\
\phantom{Fairall 9 }(Late) & & & 58530.3--59259.1 & 728.76 & 2.21 & Ed24 \\
Mrk 509 & $8.05^{+0.04}_{-0.04}$ & 0.05 & 57830.4--58103.0 & 272.63 & 1.18 & Ed19 \\
NGC 5548 & $7.69^{+0.02}_{-0.02}$ & 0.05 & 56709.6--56830.4 & 120.86 & 0.84 & Ed15\\
Mrk 817 (Epoch 1) & $7.59^{+0.06}_{-0.07}$ & 0.20 & 59176.5--59316.6 & 139.61 & 1.26 & Ca23 \\
\phantom{Mrk 817 }(Epoch 2) & & & 59316.9--59456.9 & 139.99 & 1.14 & Ca23 \\
NGC 4151 & $7.37^{+0.03}_{-0.03}$ & 0.01 & 57438.5--57507.8 & 69.24 & 0.23 & Ed17 \\
Mrk 110 & $7.29^{+0.10}_{-0.10}$ & 0.65 & 58053.6--58144.8 & 90.81 & 0.51 & Vi21 \\
Mrk 335 & $7.23^{+0.04}_{-0.04}$ & 0.07 & 58770.1--58864.3 & 92.97 & 0.44 & Ka23 \\
NGC 4593 & $6.91^{+0.07}_{-0.07}$ & 0.08 & 57583.4--57605.9 & 22.47 & 0.15 & Mc18 \\
Mrk 142 & $6.29^{+0.09}_{-0.10}$ & 3.40 & 58484.9--58604.4 & 119.59 & 0.75 & Ca20 \\
\enddata
\tablecomments{Column 1: Object. Column 2: Black hole mass with uncertainties from the AGN Black Hole Mass Database \citep{Bentz_2015}. Column 3: Accretion rate in Eddington units ($\dot{m} = L_{bol}/L_{Edd}$). Column 4: Campaign date range (MJD). Column 5: Duration of the campaign in days. Column 6: Mean sampling interval in days. For Mrk~817 and Fairall~9, two epochs are listed due to notable changes in measured lags. Column 7: Reference for the data: HS20 = \citet{Hernandez_2020}; Ed24 = \citet{Edelson_2024}, Ed19 = \citet{Edelson_2019}, Ed15 = \citet{Edelson_2015}, Ka21 = \citet{Kara_2021}, Ed17 = \citet{Edelson_2017}, Vi21 = \citet{Vincentelli_2021}, Mc18 = \citet{McHardy_2018}, Ca20 = \citet{Cackett_2020}, Ca23 = \cite{Cackett_2023}.}
\end{deluxetable*}

The AGN in our sample span two orders of magnitude in both mass and accretion rate, with masses ranging from $\sim10^6$ to $10^8 M_\odot$ and accretion rates of $\dot{m} =  L_{bol}/L_{Edd} = 0.035 - 3.4$ (see Table~\ref{tab:source_properties}). The black hole masses and their uncertainties were obtained from the AGN Black Hole Mass Database \citep{Bentz_2015}. This diversity allows us to test how disk reverberation lags scale across different physical regimes. Most of the sources accrete at relatively low rates ($\dot{m}<0.1$), with a median accretion rate of $\dot{m}=0.07$. 

The accretion rates were primarily drawn from the source-specific campaign papers listed in Table~\ref{tab:source_properties}. Most of these values were derived from SED modeling—for example, Mrk~335 from \citet{Tripathi_2020} and Fairall~9 from \citet{Vasudevan_2009}. Mrk~142 is the only super-Eddington source in the sample, with $\dot{m} = 3.4$ estimated by \citet{Cackett_2020} based on the slim disk framework of \citet{Mineshige_2000}. This is consistent with the lower limit of $\dot{m} = 2.3$ reported by \citet{Du_2018}. Mrk~142 is a clear outlier, with an accretion rate more than an order of magnitude higher than any other source in the sample; the next highest is Mrk~110 at $\dot{m} = 0.65$, based on SED fitting from \citet{Ezhikode_2017}.

It is important to note that a variety of methods were used to derive accretion rates across the sample, including broadband SED modeling and bolometric corrections. To assess the robustness of our results, we show in the appendix that our main conclusions remain generally consistent when all of the accretion rates are instead computed using a consistent method, namely the luminosity-dependent X-ray bolometric correction from \citet{Duras_2020}.

All nine AGN were targeted in recent high-cadence, multi-wavelength reverberation mapping campaigns by the Neil Gehrels Swift Observatory \citep{Gehrels_2004}. Swift monitored each object in the X-ray (0.3-10~keV) and UV/optical (1928-5468~\AA), typically at sampling rates 2-3 times faster than those of earlier campaigns. The cadence and campaign duration for each object were selected based on several factors including variability timescales informed by black hole mass as well as luminosity and practical observing constraints. Lower-mass sources like NGC~4593 and NGC~4151 were monitored intensively over shorter periods \citep[22 and 69 days, respectively;][]{McHardy_2018, Edelson_2017}, with high-cadence sampling every few hours to resolve relatively shorter timescale interband lags, whereas more massive AGN such as Mrk~509 and Fairall~9 were observed for over 270 and 700 days \citep{Edelson_2019, Hernandez_2020}, respectively, with cadences tuned to detect longer-timescale variability arising from their higher masses and thus larger disk sizes. The parameters of the sources and data used for each campaign are listed in Table~\ref{tab:source_properties}.

The Swift/UVOT data reduction largely follows the procedure detailed in the respective campaign papers in Table~\ref{tab:source_properties}. We refer the reader to these references for additional details on the UVOT data reduction procedures. Briefly, the general process for the UVOT data: processed using standard \texttt{HEASoft FTOOLS}. Source fluxes were extracted using circular apertures with background estimated from surrounding annuli, and corrections were applied for aperture losses, detector sensitivity variations, etc.

Swift X-ray light curves and spectra were produced using the Swift X-ray Telescope (XRT) product generator \footnote{\url{https://www.swift.ac.uk/user_objects/index.php}} \citep{Evans_2007, Evans_2009}. The background subtracted count rates from 0.3-10~keV were generated with a per-observation binning. We also used photometric data from the ground-based monitoring coordinated for four of the AGN--Mrk~335, Mrk~817, NGC~5548, and Mrk~110--extending spectral coverage out to the infrared. These light curves were publically available and intercalibrated across ground-based instruments using the Bayesian intercalibration method of \cite{Li_2014}, which fits a damped random walk model to each set of observations to account for differences in bandpass and sensitivity.

\begin{figure*}[t!]
    \centering
    \includegraphics[width=0.72\textwidth]{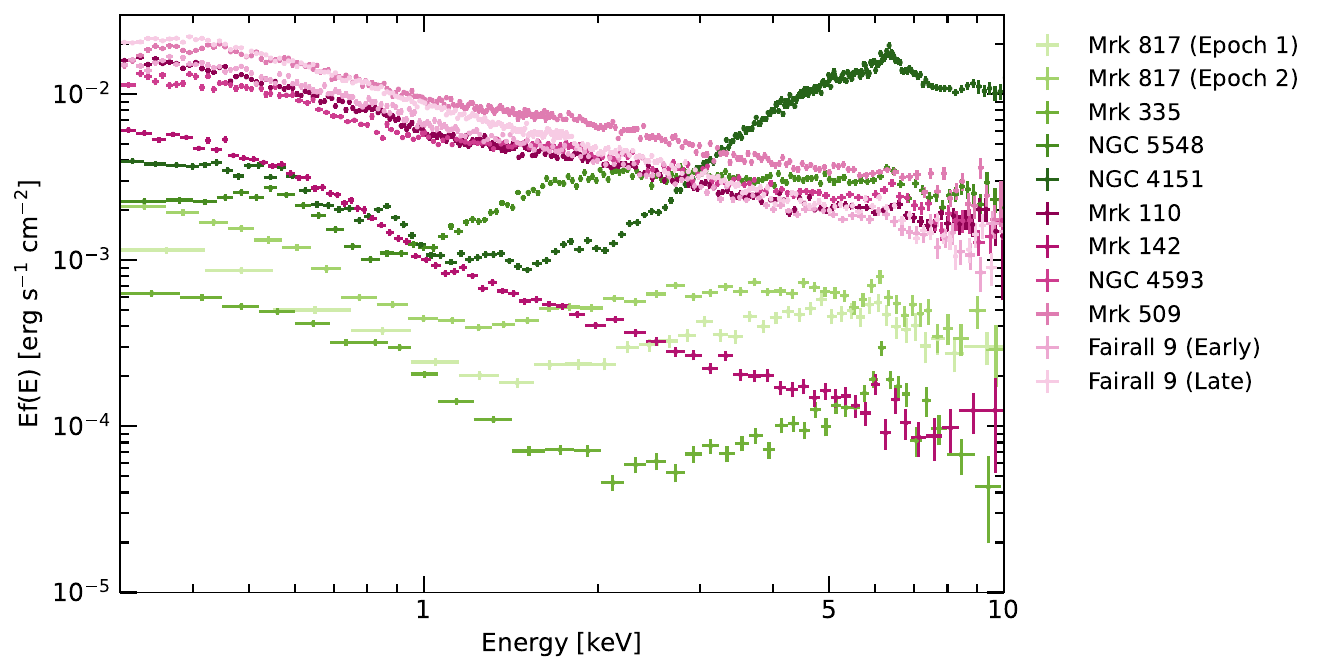}
    \caption{Swift X-ray spectra for the AGN in our sample. The sources fall into two broad categories based on their spectral shapes, which reflect varying levels of obscuration. Spectra shown in pink correspond to unobscured sources, as shown by smooth, power-law–like continua with relatively minimal spectral features. In contrast, the green spectra exhibit pronounced curvature, particularly between $\sim$1–4 keV, indicative of higher levels of obscuration from neutral and ionized gas along the line of sight. These visually apparent differences are later formalized using a Gaussian mixture model, whose classifications are consistent with this initial inspection. For detailed inspection, per-source spectra are shown in the Appendix (Figure~\ref{fig:spectra_grid}).}
    \label{fig:spectra}
\end{figure*}

The Swift X-ray spectra, shown in Figure~\ref{fig:spectra}, reveal significant diversity in spectral variables and shapes across the sample, reflecting a range of circumnuclear environments. Nonetheless, the spectra generally exhibit two general behaviors that depend on whether there is significant curvature and absorption. Five AGN--Mrk~142, NGC~4593, Mrk~110, Mrk~509, and Fairall~9--display relatively smooth, power-law-like spectra, as expected from the intrinsic coronal emission with minimal absorption (shown in shades of pink in Figure~\ref{fig:spectra}). The remaining sources--NGC~4151, NGC~5548, Mrk~335, and Mrk~817--however, are more heavily obscured, showing strong attenuation and curvature (shown in green in Figure~\ref{fig:spectra}). NGC~4151 and NGC~5548 host long-established, multi-phase warm absorbers and ionized outflows, which imprint complex absorption and emission lines and heavily absorb the flux below 2~keV \citep{Kraemer_2005, Steenbrugge_2005, Kaastra_2014, Mehdipour_2016}. The lower soft-band fluxes in Mrk~335 and Mrk~817 are instead more recent, as both sources underwent obscuration events by ionized disk winds at the time of their campaigns \citep{Longinotti_2013, Parker_2019, Zaidouni_2024}. 

In Section~\ref{subsec:classification}, we show that the distribution of spectral parameters across the sample does indeed form two statistically significant, distinct classes from which we divide the sample into two subgroups: “unobscured” and “obscured” sources (see Section~\ref{subsec:classification}).

Across both subgroups, reflection features are common: narrow Fe~K$\alpha$ emission lines are detected in nearly all sources. The spectra exhibit soft X-ray excesses below 2~keV, previously modeled as fluorescent lines produced from close to the black hole that have been relativistically smeared, or as Comptonization by a ``warm corona" in the inner disk \citep{Petrucci_2020, Mehdipour_2016}.

\section{Results} \label{sec:results}
\subsection{Spectral Modeling and $N_H$ Determination} \label{subsec:spectral_modeling}

\begin{figure}[t!]
    \includegraphics[width=\columnwidth]{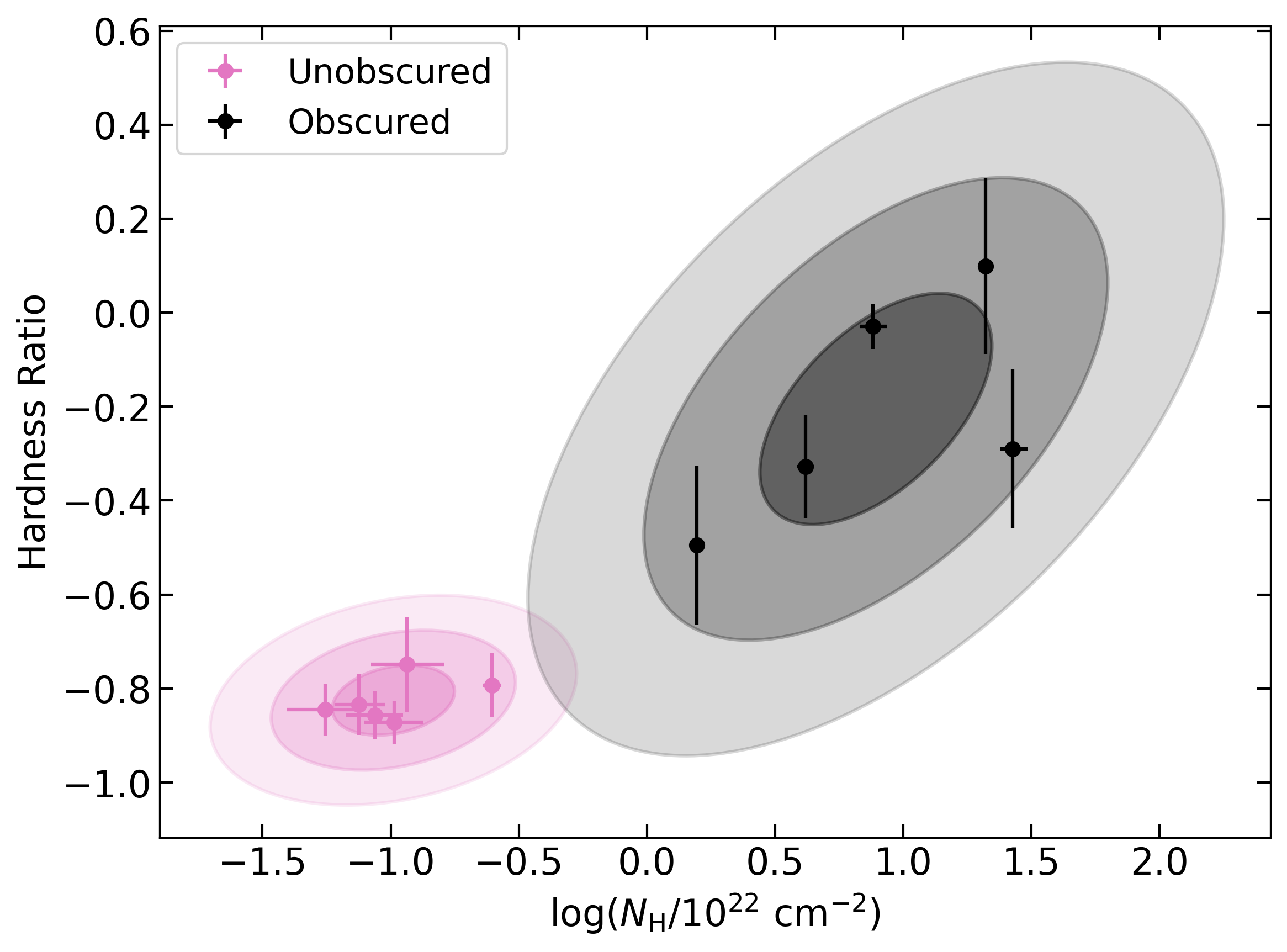}
    \caption{The two Gaussian components of the Gaussian mixture model, corresponding to the obscured and unobscured subgroups. Each distribution was fit by maximizing the likelihood of the observed column density and hardness ratio values, incorporating measurement uncertainties via a Monte Carlo approach. Covariance between the variables is accounted for, producing tilted ellipses, most notable in the obscured subgroup. The overlap between the two components is minimal, resulting in very high cluster assignment probabilities ($>99\%$) for all sources.}
    \label{fig:clustering}
\end{figure}

The UVOIR lags measured during the AGN~STORM~2 campaign of Mrk~817 showed dramatic changes coincident with changes in X-ray column density \citep{Homayouni_2023, Lewin_2024}. In light of these results, we were interested in checking for similar correlations between obscuration and the degree to which the lags exceed their standard disk predictions across a sample of AGN. We performed a simple modeling of the Swift X-ray spectra in order to constrain the column densities and thus the degree of obscuration by material along the line-of-sight. 

The spectra were fit using the model \texttt{tbabs*(ztbabs*(powerlaw+zbbody)+zgauss))}, where \texttt{tbabs} accounts for absorption within our galaxy, \texttt{ztbabs} models intrinsic absorption by additional neutral gas within the host galaxy, \texttt{powerlaw} models the primary X-ray continuum, \texttt{zbbody} is a low-temperature blackbody ($kT_e\lesssim100~\text{keV}$) accounting for the soft excess emission, and we use a phenomenological Gaussian with \texttt{zgauss} to model unresolved narrow ($\sigma = 10^{-5}~\text{keV}$) to account for Fe K emission from neutral, distant material. This initial approach using \texttt{ztbabs} was effective for the six spectra with a simple, power-law shape characteristic of unobscured or mildly absorbed AGN (i.e., the pink spectra in Figure~\ref{fig:spectra}). The final fits for these spectra using this model resulted in an aggregated reduced chi-squared of $\chi^2_\nu = 3520/3229 = 1.09$.

The remaining five spectra, belonging to Mrk~817 (both epochs), Mrk~335, NGC~5548, and NGC~4151, exhibit significant absorption that is poorly described by the model ($\chi^2_\nu = 13651/2238 = 6.11$). The \texttt{ztbabs} component assumes uniform coverage of the X-ray source, whereas these systems have been seen to have significant, complex absorption components produced by inhomogeneous and variable absorbers \citep[e.g., clumpy disk winds in NGC~5548 and Mrk~817;][]{Dehghanian_2019b, Partington_2023, Zaidouni_2024}. As such, the model shows substantial improvement when replacing \texttt{ztbabs} with \texttt{zpcfabs}, a partial covering model, likely to be more akin to the environments observed in these systems. The final model used for these four sources--\texttt{tbabs*(zpcfabs*(powerlaw+zbbody)+zgauss))}--results in a reduced chi-squared of $\chi^2_\nu = 2913/2233 = 1.30$. The fit parameters are included in the appendix.

\subsection{Identifying Obscured and Unobscured Subgroups from X-ray Spectra}
\label{subsec:classification}

As introduced in Section~\ref{sec:intro}, the AGN in our sample exhibit two broadly distinct X-ray spectral morphologies: (1) smooth, power-law-like continua characteristic of unobscured coronal emission, and (2) spectra with pronounced curvature and strong soft X-ray attenuation. The best-fit column densities support the interpretation that these behaviors reflect different levels of obscuration along the line of sight. We sought to divide the sample into ``obscured" and ``unobscured" groups to determine whether the degree of disagreement between the measured UVOIR reverberation lags and standard disk predictions differs between these two populations.

The two main shapes of the X-ray spectra in our sample motivated us to divide the AGN in our sample into two subgroups. Instead of relying on visual inspection or arbitrarily picking a column density decision boundary, we employed a \textit{Gaussian mixture model (GMM)}, which models the joint distribution of the observed variables (in our case column density and hardness ratio) as a weighted sum of two Gaussian distributions, or \textit{components}, corresponding to the two subgroups. Each component in a GMM is defined by its own mean and covariance matrix corresponding to distinct subpopulations within the data.

We model the probability of observing a data point $\mathbf{x}_i$ as a weighted sum of two Gaussian distributions, where the weights correspond to the probability of each data point $\mathbf{x}_i$ belonging to each component:
\begin{equation}
p(\mathbf{x}_i) = p_i \, \mathcal{N}(\mathbf{x}_i \mid \mathbf{\mu}_1, \mathbf{\Sigma}_1) + (1 - p_i) \, \mathcal{N}(\mathbf{x}_i \mid \mathbf{\mu}_2, \mathbf{\Sigma}_2),
\end{equation}
where $p_i$ is the probability that observation $i$ belongs to the first component (e.g., ``obscured"), $\mathbf{\mu}_k$ and $\mathbf{\Sigma}_k$ are the mean vector and covariance matrix of component $k$. We allow each component to have its own covariance matrix, so that the model can capture the variability and correlation between variables within each subpopulation.

The parameters $\{\mathbf{\mu}_k, \mathbf{\Sigma}_k\}$ are iteratively optimized by maximizing the likelihood of the observed data. An advantage of using a GMM is that it assigns each observation a probability of belonging to each component, computed after model optimization using Bayes’ rule. These probabilities reflect the confidence of each classification, i.e., how strongly a given data point is associated with each subpopulation. 

To propagate measurement uncertainties through the modeling process, we implemented a Monte Carlo resampling approach: we generated $10^4$ synthetic datasets by resampling the data from Gaussian distributions centered at their observed values and widths set by the individual measurement errors. A GMM model was then fit to each synthetic dataset. The distribution of model parameters across realizations were then used to constrain final estimates and standard errors.

To improve the robustness of the clustering, we extended the GMM analysis to two dimensions by including the hardness ratio, $HR = (H-S)/(H+S)$, where $S = \text{Flux}(0.5$–$2~\text{keV})$ and $H = \text{Flux}(4$–$7~\text{keV})$. The hardness ratio serves as a less model-dependent proxy for column density, and so the two parameters are likely not independent. To account for this, covariance is a free parameter optimized in both component's covariance matrix. In any case, the two-dimensional clustering results are generally consistent with those from fitting a one-dimensional GMM to either column density or hardness ratio alone.

The results of the GMM classification are shown in Figure~\ref{fig:clustering}, which align well with our initial visual assessments. Component 1, corresponding to the unobscured AGN, is centered at the best-fit mean of $N_{\mathrm{H}} = (1.02 \pm 0.14) \times 10^{21}~\mathrm{cm}^{-2}$ and $HR= -0.82\pm0.03$. This distribution is relatively compact, and shows low dependence between column density and hardness ratio. In contrast, component 2, corresponding to the obscured AGN, is centered at a higher mean $N_{\mathrm{H}} = (7.76 \pm 0.72) \times 10^{22}~\mathrm{cm}^{-2}$ and $HR = -0.21\pm0.07$. The obscured distribution exhibits a broader, more elliptical shape, driven by larger variances in both covariates and a stronger covariance between them.

The motivation to formally split the sample is well-supported by the clustering outcome: the obscured and unobscured components are well-separated beyond the $2\sigma$ level in the joint parameter space, with only $\sim 0.7\%$ overlap in parameter space\footnote{We estimated the overlap between the two Gaussian components using the Bhattacharyya coefficient ($BC$), finding $BC \approx 0.007$.} As a result, each AGN resulted in a very high probability ($99.7\%$ on average) of belonging to its assigned group, indicating strong confidence in the assignments by the model. 

While Mrk~142 is classified in the unobscured subgroup based on its spectral properties, we treat it as a unique case, belonging to a third class: ``Super-Eddington." The source's exceptionally high accretion rate ($\dot{m} = 3.4$) lies two orders of magnitude above the median accretion rate of the sample and is the only source in the super-Eddington regime. To ensure that this manual reassignment did not affect the separation between the obscured and unobscured populations, we repeated the GMM clustering excluding Mrk~142 and found that the results remained consistent. 

\subsection{Fitting UVOIR Lags Across the Sample}

\begin{figure*}[t!]
    \centering
    \includegraphics[width=\linewidth]{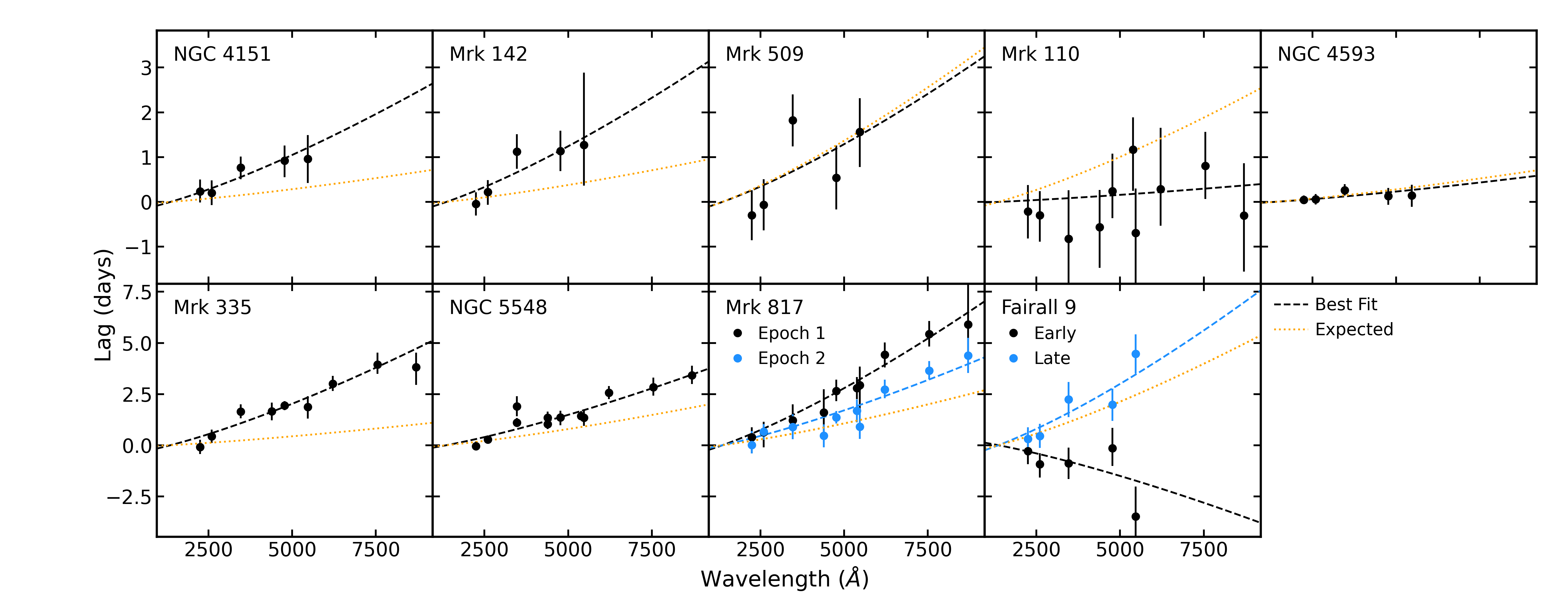}
    \caption{Measured interband lags as a function of wavelength using the ICCF approach. The dashed black line shows the best-fit power-law relation, $\tau \propto \lambda^{4/3}$, with the normalization treated as a free parameter. The orange dotted line shows the same relation, but with the normalization instead predicted from standard accretion disk given the black hole mass and accretion rate of each source. As such, the differences between these measured and predicted curves reflect departures from theoretical expectations (the ``normalization excess").}
    \label{fig:lags_vs_wavelength}
\end{figure*}

\begin{figure}[t!]
    \centering
    \includegraphics[width=\columnwidth]{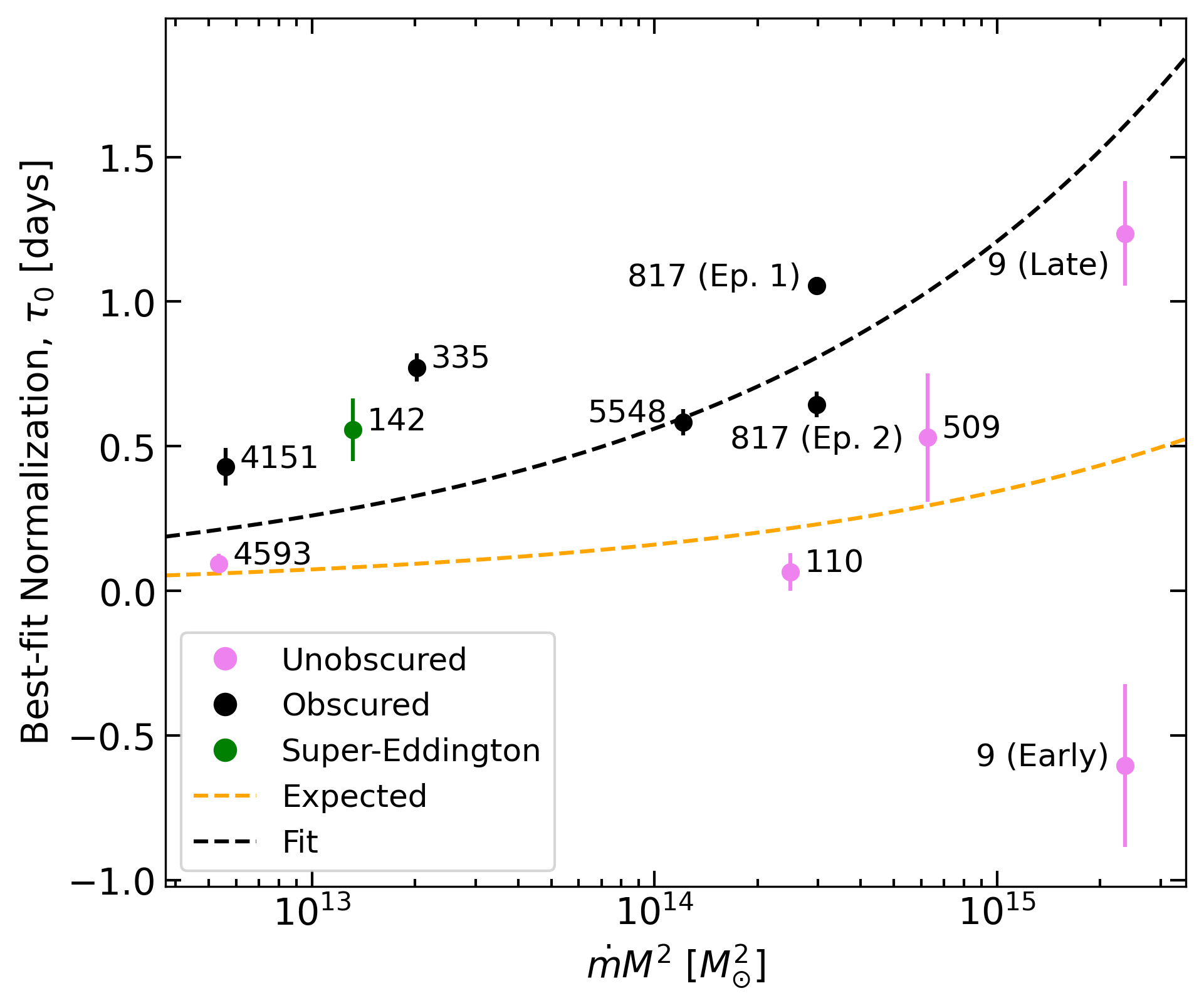}
    \caption{Best-fit lag normalizations ($\tau_0$), corresponding to the dashed black curves in Figure~\ref{fig:lags_vs_wavelength}, plotted against $M^2 \dot{m}$. The orange dashed line shows the predicted scaling $\tau_0 \propto (M^2 \dot{m})^{1/3}$, computed from a modified version of Equation 12 in \citet{Fausnaugh_2016}, assuming standard disk parameters stated in the text. Most sources lie above the theoretical prediction, highlighting a systematic excess in lag normalizations relative to expectations. However, while the obscured sources are well described by the best-fit relation, the unobscured sources tend to lie closer to the predicted scaling, suggesting that the excess lag normalizations are most significantly driven by the obscured population.}
    \label{fig:norms_vs_massmdot}
\end{figure}

\begin{figure*}[t!]
    \centering
    \includegraphics[width=0.5\textwidth]{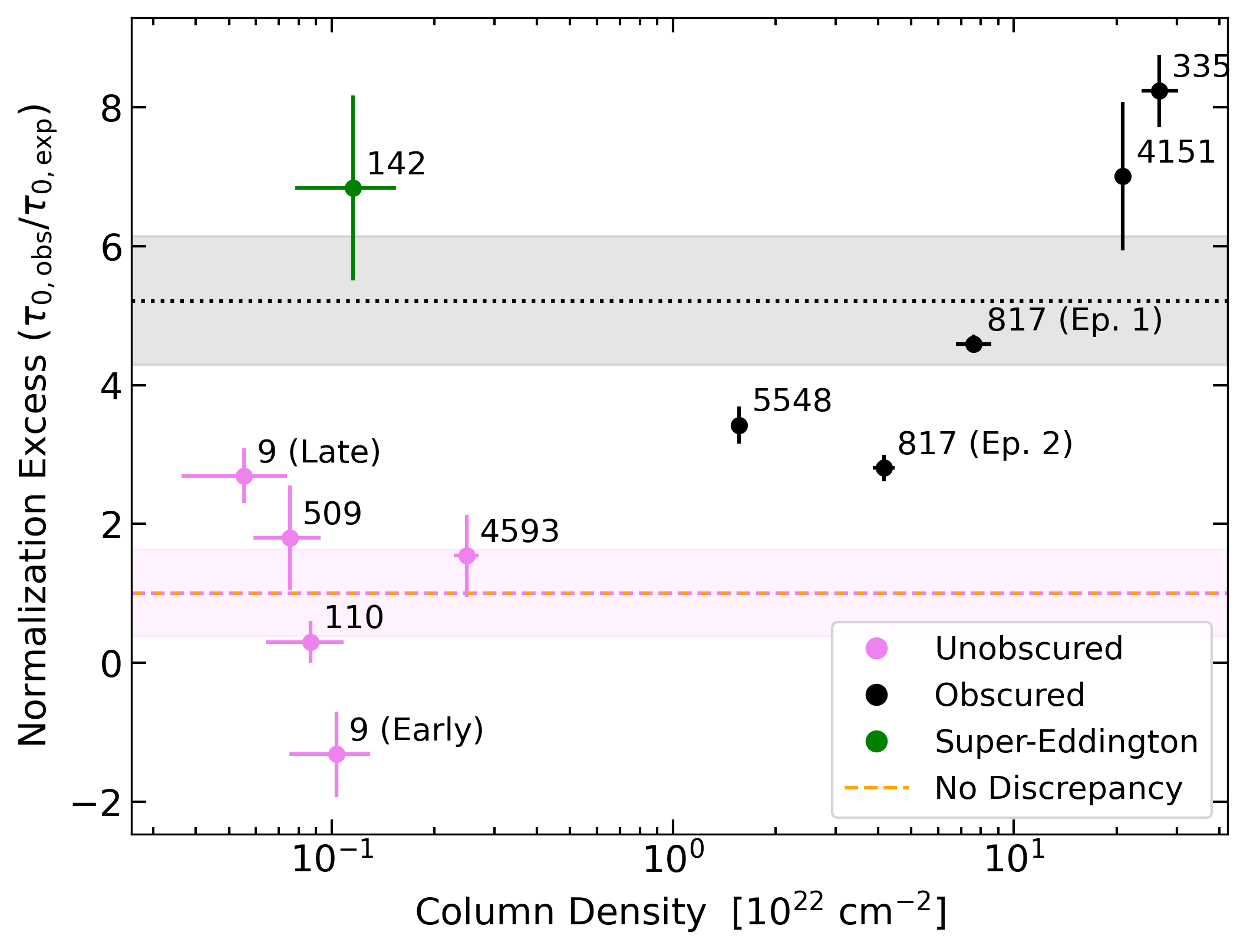}
    \caption{Normalization excesses--defined as the ratio of the measured lag normalization to the predicted value--for each AGN in the sample. Dotted lines indicate the sample means for each unobscured and obscured subgroup, with shaded bands denoting the standard error of the mean. The full sample has a mean excess of $3.45 \pm 0.27$, with the obscured subgroup showing a larger average excess ($5.21 \pm 0.47$) than the unobscured group ($1.00 \pm 0.31$). A similar trend is observed in the U-band lag excesses on their own. Hypothesis testing confirms that the excesses in the full and obscured samples are significantly greater than one ($p = 0.011$ and $p = 0.008$, respectively), while the unobscured group (as expected) shows no significant deviation from theoretical expectations ($p = 0.50$). The one super-Eddington source (Mrk~142) occupies a distinct region of the parameter space.}
    \label{fig:norms_vs_nh}
\end{figure*}

We measure the time lags between the light curve in each wave band with respect to the UVW2 reference band using the Interpolated Cross-Correlation Function (ICCF) method of \citet{Peterson_1998}. This widely used approach in reverberation mapping ensures consistency with previous campaign studies \citep[e.g.,][]{Edelson_2019, Fausnaugh_2016, Kara_2021}. In summary, the ICCF shifts one of the light curves and determines the correlation coefficient by linearly interpolating the other. This approach measures time lags present due to variability on roughly the longest timescales \citep[i.e., roughly consistent with the lowest-frequency lags in the Fourier-resolved approach;][]{Lewin_2023}. 

The lag uncertainties were estimated by using a Monte Carlo approach known as flux randomization/random subset selection (FR/RSS): $10^4$ realizations of the original light curves with a random subset of the original data points are generated, with each flux measurement redrawn from a normal distribution defined by the observed flux and its uncertainty. The centroid lag across realizations produces a distribution from which the median and 16th and 84th quantiles are used to estimate the final lags and their uncertainties.

This procedure was executed using STELA Toolkit \footnote{STELA Toolkit \url{https://github.com/collinlewin/STELA-Toolkit}}, a Python package developed by the lead author. While STELA includes tools for computing time-domain lag measurements such as the cross-correlation function (CCF) used for the results of this paper, its primary functionality is built around Gaussian Process interpolation of light curves to enable measuring data products in the frequency domain. These include the power spectral density, time lags, coherence, etc. The package is fully documented and beginner-friendly,  including a tutorial.

The measured UV/optical reverberation lags are compared in Figure~\ref{fig:lags_vs_wavelength} to those predicted from standard accretion disk theory as a function of wavelength:
\begin{equation}
\tau(\lambda) = \tau_0 \left[ \left( \frac{\lambda}{\lambda_0} \right)^{4/3} - 1 \right],
\end{equation}
where $\tau_0$ is the \textit{lag normalization}. The normalization is estimated for each source via a least-squares fitting to the measured lags.

The normalization $\tau_0$ is expected to scale with black hole mass and accretion rate as $\tau_0 \propto (M^2 \dot{m})^{1/3}$, where $\dot{m} = L_{\mathrm{Bol}} / L_{\mathrm{Edd}}$ is the Eddington ratio \citep[via a simple modification of equation 12 from][]{Fausnaugh_2016}. Since mass and accretion rate vary across the sample, this scaling provides a convenient way to compare the lag amplitudes between sources using a single parameter ($M^2 \dot{m}$). The proportionality constant of this relation $\alpha$ depends on physical constants and parameters of the accretion disk, for which we assume standard values of $\eta = 0.1$ for the radiative efficiency, $\kappa = 1$ for the ratio of external to internal heating, and $X=2.49$ for the temperature-wavelength conversion correction using flux-weighted mean radius \citep{Fausnaugh_2016}. 

The estimated normalizations are plotted against $M^2 \dot{m}$ in Figure~\ref{fig:norms_vs_massmdot}, with the expected values from standard disk reprocessing shown in orange. All but two measured normalizations--Mrk~110, Fairall~9 during the first 300 days of its campaign--exceed their predicted values. The normalizations show a large amount of scatter across $M^2 \dot{m}$-space, suggesting a complex dependence on parameters beyond mass and accretion rate alone.

For the two sources with the longest campaigns, Mrk~817 and Fairall~9, the large amount of data (relative to the source-dependent timescale of reprocessing) enabled lag measurements in multiple time intervals, across which the measured lags change. For Mrk~817, we used epochs~1 and 2 from \citet{Lewin_2024}, corresponding to periods of high and low column density, across which the lag amplitudes varied by a factor of two with a highly significant change in lag normalization. For Fairall~9, we measured the lags before and after $\text{MJD}=58529.2$, i.e., the end of the campaign window used by \citet{Hernandez_2020}. This division results in significantly different lag behaviors that span the full range of best-fit normalizations in our sample: $\tau_0 = -0.6-1.24~\text{days}$. During the first $\sim$300~days of the campaign, we find a negative lag normalization of $\tau_0=-0.6~\text{days}$. Our time range closely follows that of \citet{Hernandez_2020}, except that we include the first 20 days—which they excluded to align with ground-based coverage—since our analysis relies exclusively on Swift data. Although their initial analysis did not report a negative normalization, they recovered a similar result when isolating variability on the longest timescales, consistent with the behavior we observe by using the full Swift baseline in this window. \citet{Hagen_sub} found similarly significant changes in lag behavior during the second half of the monitoring period.

If we instead fit the measured normalizations for the proportionality constant $\alpha$, we find a best-fit value of $\alpha = 3.4\times10^{-5}~\text{days}~M_\odot^{-2/3}$. This is roughly three times larger than the expected value of $\alpha = 1.2\times10^{-5}~\text{days}~M_\odot^{-2/3}$, assuming the typical aforementioned values for parameters $\kappa$ and $\eta$. In other words, we recover the accretion disk size problem in this uniform sample. It is difficult to reconcile this with model parameter assumptions alone: increasing $\kappa$ by an order of magnitude causes $\alpha$ to increase only by a factor of 1.5, since $\alpha \propto \left( \frac{\kappa}{\eta} \right)^{1/3}$. Instead, adopting a larger value of $X = 5.04$, as used in \citet{Tie_2018}, reduces the predicted lags and can partially mitigate the discrepancy, though it does not fully resolve the excess observed across the sample.Similarly, the measured masses would need to be $\sim27$ times greater on average to meet the expected normalizations given the $\alpha\propto M^{1/3}$-scaling \citep{Fausnaugh_2016}. Such a substantial change in mass would be especially difficult to reconcile via systematic uncertainties in the BLR geometry and inclination (and thus the virial factor).

All but one of the sources below the best-fit relation are unobscured; in fact, the theoretical prediction provides a better overall fit for the unobscured sources, with a reduced chi-squared of $\chi^2_\nu = 1.47$. This motivates the idea that the longer-than-expected lag normalizations are significant only for the subset of obscured AGN, which we explore more rigorously in the next subsection.

\subsection{Testing for Lag Excesses: Obscured vs. Unobscured AGN}

To quantify the extent to which the observed lag normalizations deviate from the standard disk prediction, we define the \textit{normalization excess} as the ratio of the measured lag normalization to the theoretical value, where the theoretical normalization is computed for each source based on its black hole mass and accretion rate. As shown in Figure~\ref{fig:norms_vs_nh}, the normalization excesses span a wide range across the sample, from $-1.32$ (Fairall~9, early campaign) to $8.24$ (Mrk~335). The mean excess is $3.45 \pm 0.27$, with a standard deviation of 2.97, indicating that, on average, the observed UV/optical lags are more than three times longer than predicted by standard accretion disk theory, for our fiducial parameters ($X=2.49, \kappa=\eta=0.1$. There is large scatter as a result of the significant difference in excesses between the unobscured and obscured subgroups.

While these values provide a useful summary of the broader sample, we saw in the previous subsection (Figure~\ref{fig:norms_vs_massmdot}) that much of the tension with thin-disk theory appears to be driven by the obscured sources. When we examine the normalization excesses within the obscured and unobscured subgroups, we find a similar pattern: unobscured sources exhibit much smaller excesses than their obscured counterparts on average. In fact, the mean excess in the unobscured group is precisely $1.00 \pm 0.31$, indicating that the unobscured subgroup does show statistical consistency with theoretical expectations. In contrast, the obscured group shows a substantially higher mean excess of $5.21 \pm 0.47$. In the case of Mrk 817, where we observed significant variability in $N_{\rm H}$ over the course of the campaign, the lower $N_{\rm H}$ epoch yields a lag normalization much closer to the expected value (see \citealt{Lewin_2024} for details).

While sample statistics provide useful estimates, formally inferring properties of the underlying population requires hypothesis testing. To assess the significance of the normalization excesses, we performed one-sided $t$-tests at the $\alpha = 0.05$ significance level. We tested for normality to verify that the assumptions underlying the $t$-test were met using Shapiro-Wilk and Anderson-Darling tests. The Shapiro-Wilk test yielded a test statistic of 0.96 with a $p$-value of 0.80, while the Anderson–Darling test statistic was 0.24, far from the critical value at even the 15\% significance level. We also examined a quantile-quantile plot, which showed no notable visual deviations from normality.

The null hypothesis for the tests was $H_0: \mu = 1.0$, corresponding to lags consistent with standard disk theory, with the alternative hypothesis of $H_1: \mu > 1.0$, indicating longer-than-expected lags. For the full sample, we find strong evidence that the average lag normalization exceeds thin-disk predictions ($p = 0.011$), with this result driven primarily by the obscured subgroup, which on its own shows even stronger significance ($p = 0.008$). In contrast, the unobscured group yields a $p$-value of $p=0.50$, indicating little to no evidence against the null hypothesis; that is, their lags are statistically consistent with the theoretical expectation. The same conclusions also hold when repeating these tests using only a single epoch for Fairall~9 and Mrk~817, confirming that the significance is not artificially inflated by including potentially dependent measurements from multiple epochs of the same source. Likewise, if we exclude the early portion of the Fairall~9 campaign---during which the lags decrease with wavelength and are therefore unlikely to reflect canonical reverberation, making a $\tau \propto \lambda^{4/3}$ fit not necessarily appropriate---we still find the same qualitative conclusions, though the overall significance weakens slightly, yielding a $p$-value of 0.16.

\begin{figure}[t!]
    \centering
    \includegraphics[width=\columnwidth]{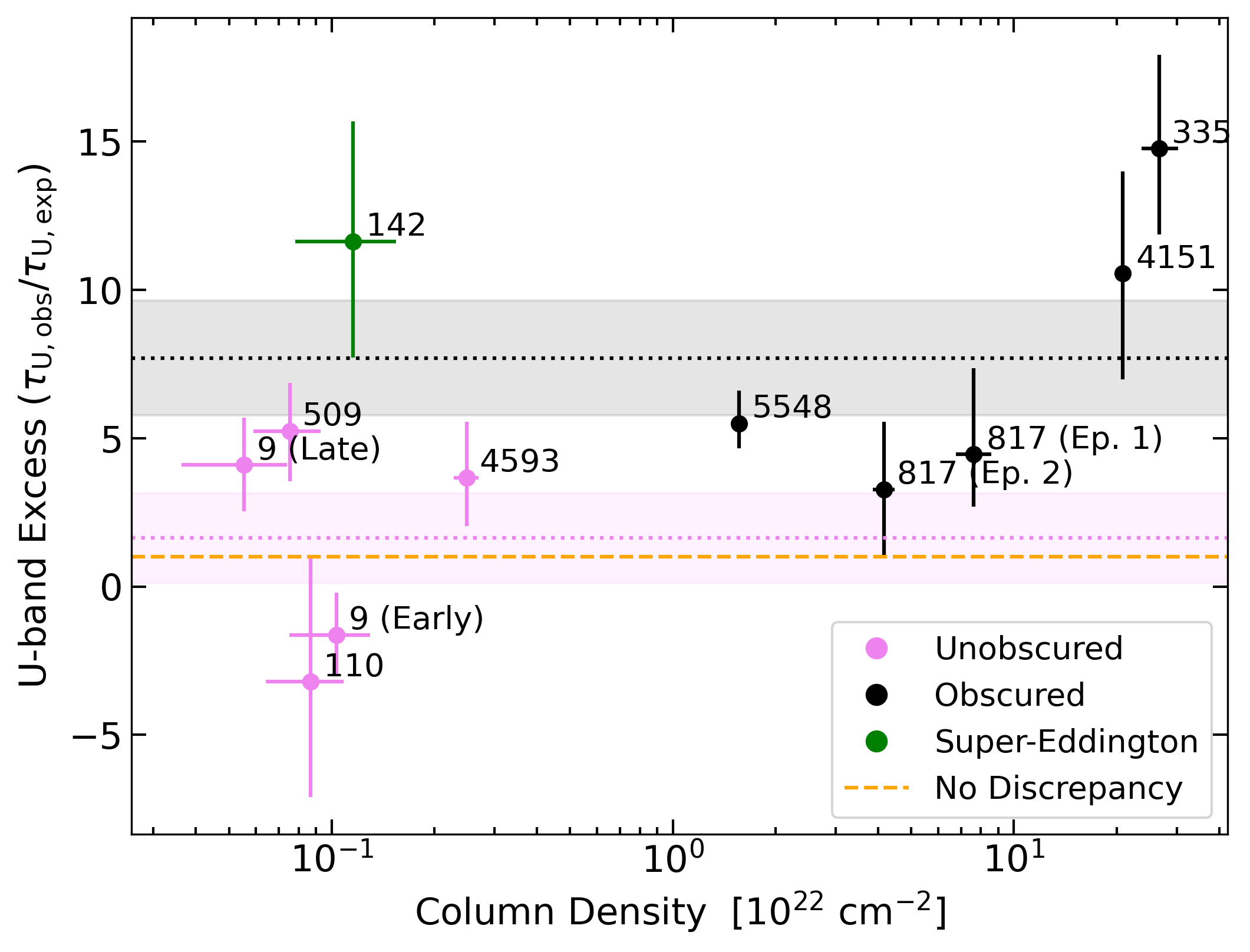}
    \caption{U-band excess as a function of X-ray column density. Dotted lines indicate the sample means for each unobscured and obscured subgroup, with shaded bands denoting the standard error of the mean. The mean excess across the full sample is $5.30 \pm 0.49$, $7.70 \pm 0.96$ in the obscured subgroup, and $1.63 \pm 0.76$ in the unobscured subgroup. Like the full normalization excess, only the unobscured subgroup reveals no significant deviation from theoretical expectations ($p = 0.012$ for the obscured group vs. $p = 0.36$ for the unobscured group).}
    \label{fig:U_excess_vs_nh}
\end{figure}

We can also consider the lag excesses in the U band (the ratio of the observed-to-predicted lag in just this band). As shown in Figure~\ref{fig:U_excess_vs_nh}, the mean U-band lag excess is $5.30 \pm 0.49$ for the full sample, $7.70 \pm 0.96$ for the obscured subsample, and $1.63 \pm 0.76$ for the unobscured subsample. Repeating the testing procedure under the same hypotheses, we again find strong evidence that the full sample exhibits longer-than-expected lags ($p = 0.018$)\footnote{Computed using a $t$-test, justified by sufficient normality (Shapiro–Wilk $p = 0.57$).}. This result is driven by the obscured subgroup ($p = 0.012$), while the unobscured sources remain consistent with the thin-disk model ($p = 0.36$).

These results suggest that, at the population level, obscured AGN exhibit significantly larger normalization excesses than unobscured AGN, which appear broadly consistent with the thin-disk prediction.

\subsection{Correlations Among Physical Parameters and Lag Excess}

In the previous section, we found statistical evidence that normalization excess is significantly larger in AGN with higher column densities, raising the possibility that other parameters may also be related to the excess. To explore this further, we turn to a broader correlation analysis.

We measured all pairwise Spearman's rank correlation coefficients ($\rho$) between the following variables: black hole mass ($M$), mass accretion rate ($\dot{m}$), normalization excess (i.e., $\tau_0$ excess), the X-ray to UV lag amplitude ($\tau_{\rm X}$), the maximum linear correlation coefficients between the X-ray and UV bands (UVW2 to 0.3–10~keV: $R_{\rm max}^{\rm X-ray}$) and between UV bands (UVW2 to V: $R_{\rm max}^{\rm UV}$), column density ($N_{\rm H}$), the optical–to–X-ray spectral slope ($\alpha_{\rm ox}$), and the fractional UVW2 root-mean-square (RMS) variability. 

The optical-to-X-ray spectral slope $\alpha_{\rm ox}$ quantifies the relative strength of X-ray to UV emission:
\begin{equation}
\alpha_{\rm ox} = 0.3838 \, \log \left( \frac{L_{\rm 2\,keV}}{L_{2500~\text{\AA}}} \right),
\end{equation}
where $L_{\rm 2 ~\rm{ keV}}$ and $L_{2500~\rm{\AA}}$ are the monochromatic luminosities at 2~keV and 2500~\AA, respectively. We estimated the unabsorbed 2 keV fluxes using our X-ray spectral fits with the \texttt{cflux} component applied to only the power-law. For the 2500~\AA ~fluxes, we first performed a linear interpolation between the Swift UVW1 and UVM2 bands to estimate the observed flux at 2500~\AA. We then corrected for Galactic extinction along the line of sight using the \texttt{gdpyc} package \citep{Ruiz_2018}, which models dust attenuation based on the source coordinates and standard extinction laws.

The fractional RMS is computed following the formulation of \citet{Vaughan_2003}.
A majority of the Eddington ratios were derived using the bolometric luminosities from the literature because in many cases, these studies do modeling of the spectral energy distribution (SED), rather than scaling. However, we show in the appendix that instead assuming the luminosity-dependent X-ray bolometric correction from \citep{Duras_2020} to estimate the Eddington ratios produces consistent results and thus conclusions.

The correlation coefficients and their uncertainties, determined from their associated 68\% confidence intervals, were estimated using a Monte Carlo approach of repeatedly resampling the covariates according to their reported measurement uncertainties. Spearman’s rank correlation computes the Pearson correlation coefficient on the ranks of the data rather than the raw values, allowing us to measure monotonic relationships, regardless of linearity. We adopted this method because we expect that many of the underlying relationships between parameters are likely nonlinear, and so relying on linear correlation alone could be misleading.

\begin{figure*}[t!]
    \centering
    \includegraphics[width=0.65\textwidth]{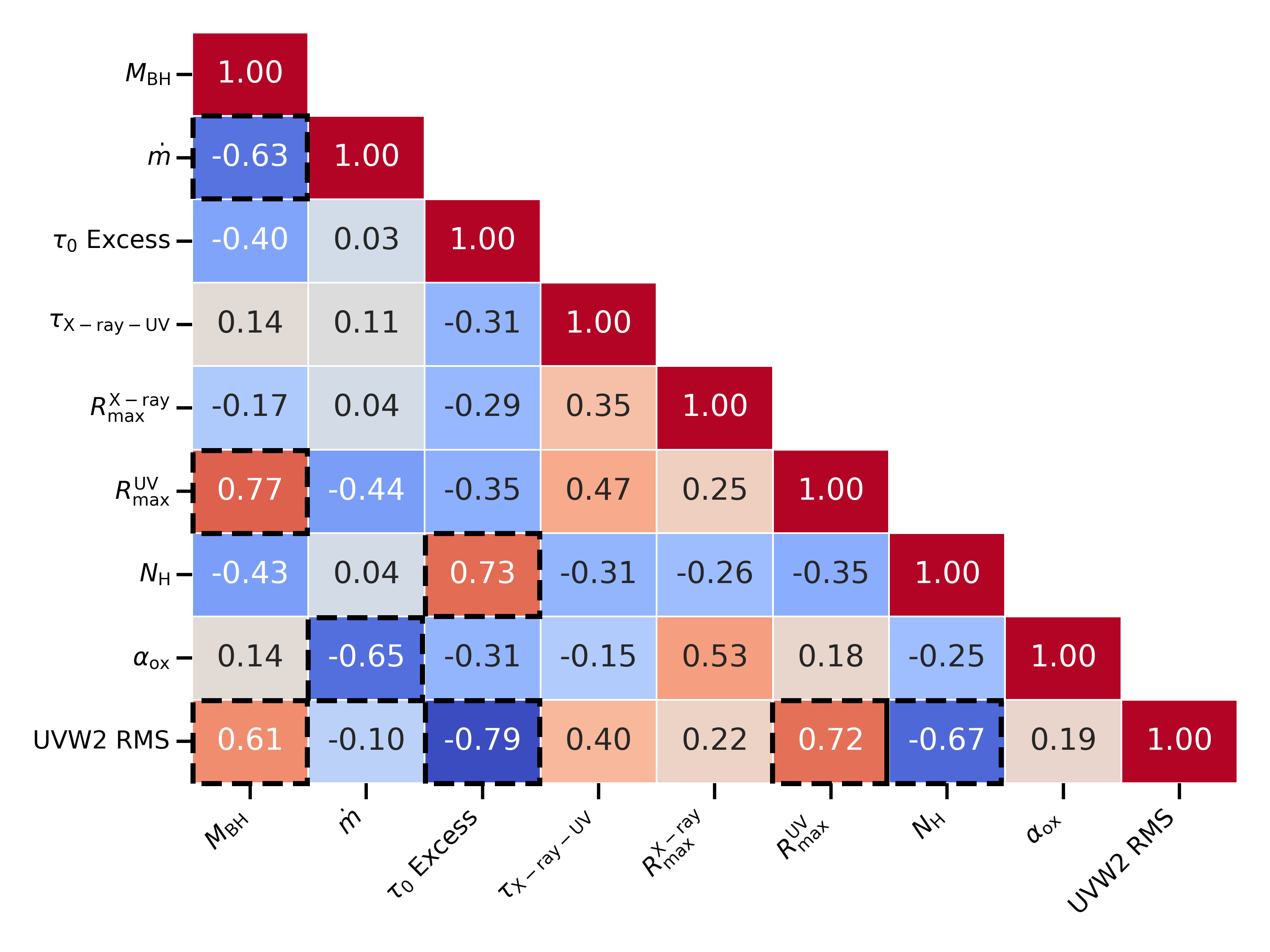}
    \caption{Spearman's rank correlation matrix among key parameters in the sample, including black hole mass, accretion rate, normalization excess, X-ray to UV lag, column density ($N_{\rm H}$), fractional UVW2 RMS variability, optical–to–X-ray spectral slope ($\alpha_{\rm ox}$), and maximum correlation coefficients ($R_{\rm max}$). In order to assess which correlations reflect real associations rather than occurring by chance, we computed $p$-values using a permutation test: one variable in each pair was randomly shuffled many times to build the distribution of correlations expected under the null hypothesis of no relationship. Black dashed outlines highlight boxes for correlations that are statistically significant at the $\alpha = 0.05$ level, corresponding to $|\rho|\geq 0.61$.
}
    \label{fig:correlations}
\end{figure*}

The resulting correlation matrix is shown in Figure~\ref{fig:correlations}. Given the sample size ($n=11$), it is relatively common to observe apparent correlations by chance. To formally assess which correlations are likely to be meaningful, we computed $p$-values for each coefficient under the null hypothesis of no correlation $H_0: \rho = 0$ vs. $H_1: \rho\neq 0$. The $p$-values were estimated using a permutation test: we randomly permuted the variables to break any correlation and computed the resulting coefficient. Repeating this $10^4$ times results in a distribution under the null hypothesis that reflects the correlations expected solely by chance in a sample of this size.

Only relatively strong correlations ($|\rho| > 0.61$) achieve statistical significance at the 0.05 level; for context, the coefficient between normalization excess and column density of $\rho = 0.73$ corresponds to a $p$-value of $p=0.014$. As a result, we focus our interpretation on the correlations that exceed this threshold, but also mention those whose uncertainties overlap with it.

There are eight significant correlations among the variables of interest. Focusing first on our primary target thus far, the normalization excess: its strongest correlations are with column density ($\rho = 0.73^{+0.10}_{-0.08}$) and the UVW2-band fractional RMS variability ($\rho = -0.79^{+0.05}_{-0.05}$). UVW2 RMS shows several correlations, including with column density ($\rho = -0.67^{+0.06}_{-0.10}$), maximum UV–optical correlation $R_{\rm max}$ ($\rho = 0.72^{+0.00}_{-0.03}$), and, marginally, black hole mass ($\rho = 0.61^{+0.05}_{-0.07}$).

These relationships are physically intuitive: for instance, high $R_{\rm max}$ and low RMS both reflect more coherent, less stochastic variability, which is often associated with more massive black holes where fluctuations are smoothed over longer timescales. This interpretation is further supported by the strong positive correlation between black hole mass and maximum UV–optical correlation $R^{\rm UV}{\rm max}$ ($\rho = 0.77^{+0.05}{-0.02}$), suggesting that higher-mass systems tend to exhibit more temporally stable, cross-band variability behavior.

The remaining notable correlations are anti-correlations between black hole mass and accretion rate ($\rho = -0.63^{+0.03}_{-0.09}$) $\alpha_{\rm OX}$ and accretion rate ($\rho = -0.65^{+0.02}_{-0.06}$). The anti-correlation between mass and accretion rate is likely a selection effect as a result of being count-limited for the campaigns; at a given redshift, higher mass objects allow us to probe lower accretion-rate sources at the same overall count rate. Also important is the variability timescale, which scales with $M \dot{m}^{-1}$ \citep{Mchardy_2006}, enforcing a trade-off between mass and accretion rate in order to measure sufficient variability during the campaigns.

We also note several relationships involving the X-ray to UV lags ($\tau_{\rm X}$) that yield correlation coefficients that are not formally significant as a result of the large uncertainties on the lags; for instance, $\tau_{\rm X}$ shows potential positive correlations with fractional UVW2 RMS ($\rho = 0.40^{+0.25}_{-0.16}$), with $R^{\rm UV}_{\rm max}$ ($\rho = 0.47^{+0.29}_{-0.00}$), and with $R^{\rm X-ray}_{\rm max}$ ($\rho = 0.35^{+0.26}_{-0.18}$).

There are overlapping relationships that suggest that caution is needed when interpreting the individual correlations in isolation. In particular, while the strong anti-correlation between RMS variability and normalization excess may appear physically meaningful on its own, it could instead reflect their shared relationships with other variables, namely column density. Given that column density is positively correlated with the normalization excess and negatively correlated with RMS, the correlation between RMS and excess may be partially or wholly driven by the obscuration traced by column density. In this case, RMS might act more as a proxy for column density than as an independent driver of the lag normalization, although we check for the latter in Section~\ref{sec:discuss}. The additional correlations between RMS and parameters like $R_{\rm max}^{\rm UV}$ and black hole mass \citep[see also, e.g.,][]{ponti_2012} further suggest that RMS variability may encode broader structural or observational effects within the AGN sample.

\subsection{Multivariate Regression: Identifying Predictors of the Lag Excess}

To address these interdependencies, we apply a multivariate linear regression to assess which variables most directly predict the normalization excess. Visual inspection of the relationships between normalization excess and each predictor suggest approximate linearity, with residuals that appear reasonably homoscedastic (constant over the input domain). Together with our earlier normality tests, this supports the use of linear regression as an appropriate modeling choice. We exclude Mrk~142 from the regression, as its status as the only super-Eddington source makes it a significant outlier that would overly influence the model, particularly by preventing the inclusion of accretion rate as a meaningful predictor.

Before performing the regression, we assessed multicollinearity among the predictors by computing the Variance Inflation Factor (VIF) for each variable. High multicollinearity can inflate the coefficients and their standard errors, which will matter for testing significance. The VIF for each predictor is computed by regressing that predictor on all other predictors in the model and assessing how well it can be linearly predicted from them. All VIF values were below the commonly used threshold of 5, indicating no severe collinearity.

Given the small size of our dataset and the relatively large number of candidate variables, the model overfits, i.e., captures noise instead of meaningful trends. To address this, we used Lasso regression, which adds a penalty on model complexity and automatically removes weak predictors by shrinking their coefficients to zero. This is done by adding a regularization term to the standard least-squares loss function:
\begin{equation}
\mathcal{L}(\beta) = \sum_{i=1}^{n} \left(y_i - \mathbf{x}_i^\top \beta\right)^2 + \lambda \sum_{j=1}^{p} |\beta_j|,
\end{equation}
where $\beta$ represents the regression coefficients, and regularization strength $\lambda$ (higher values of $\lambda$ lead to more coefficients being reduced to zero).

To choose the best value of the regularization strength $\lambda$, we used leave-one-out cross-validation, in which the model is trained on all but one data point and then tested on the held-out point. This is repeated across all observations to find the value of $\lambda$ that minimizes average prediction error (in our case, $\lambda^* = 0.53$).

As expected from the correlation analysis, this approach retains two variables: column density and fractional RMS variability.

We then performed standard (ordinary least squares) linear regression using only these two predictors, and then assessed the significance of the selected variables using hypothesis testing ($t$-tests) on the regression coefficients. The overall model fit was strong, with $R^2 = 0.84$ and an adjusted $R^2 = 0.80$, which accounts for the number of predictors in the model. The coefficient for column density was positive and statistically significant ($\beta_{N_H} = 1.96$, $p = 0.012$)\footnote{The model coefficients are reported in standardized units, as the predictor variables were scaled to have zero mean and unit variance prior to model fitting.}, while the coefficient for fractional RMS variability was not ($\beta_{\rm RMS} = -0.73$, $p = 0.247$). 

If we fit a model using an input of only column density, the performance decreases slightly, with $R^2 = 0.80$ and adjusted $R^2 = 0.78$. This level of performance is strong, especially given that the model uses only a single predictor, yet explains over 80\% of the variance in the normalization excess. Fitting the excesses using RMS alone produces a notably poorer fit ($R^2 = 0.60$).

These results indicate that, after controlling for the RMS variability, column density is a clearly strong and independent predictor of the normalization excess. In contrast, fractional RMS does not remain significant once column density is included, suggesting that its strong correlation with normalization excess may be driven by its shared covariance with column density.

\section{Discussion} \label{sec:discuss}

We have presented evidence that the accretion disk size problem in AGN disk reverberation mapping is not uniformly present across the AGN population, but is instead strongly linked to the presence of line-of-sight obscuration. While the average lag normalization is more than three times larger than predicted, once divided by spectral properties into obscured and unobscured subgroups, the discrepancy is found entirely in the obscured AGN: the unobscured sources show a mean normalization excess consistent with the standard accretion disk ($1.00 \pm 0.31$), while the obscured group shows a significantly higher mean of $5.21 \pm 0.47$ ($p=0.008)$. This conclusion holds for both the full UVOIR lag-wavelength spectrum and the U-band lag excess on its own.

In addition to a strong correlation with column density ($\rho = 0.73$), the normalization excess also shows a strong anti-correlation with UVW2-band fractional RMS variability ($\rho = -0.79$), which itself is anti-correlated with column density ($\rho = -0.67$). These results raise the question of whether obscuration influences the measured lags directly, or indirectly by suppressing variability and thereby altering the lag recovered from the CCF. In either case, the obscuring material traced by column density plays a central role.

To test whether low RMS introduces systematic effects in the ICCF lag measurements--potentially explaining the observed anticorrelation--we performed a series of simulations. Light curves were generated using the method of \citet{Timmer_Konig_1995} across 10 RMS values spanning the range observed in our sample, assuming a power spectral density (PSD) slope of $s_X = -2$. We obtained consistent results when testing slightly shallower or steeper slopes ($s_X = -1.5$ and $-2.5$), in line with empirical PSD estimates \citep{Panagiotou_2022a}. To simulate lag behavior, we injected a delay using a log-normal impulse response function designed to mimic the general shape of frequency-resolved lags observed in AGN \citep[e.g.,][]{Cackett_2022, Lewin_2023, Lewin_2024}.

Across the range of RMS values consistent with our observed sources, the recovered lags show no significant trend or systematic overestimation. Even at RMS levels well below those observed, where the CCF becomes increasingly noisy and uncertain, we find no evidence for a significant anticorrelation between RMS and the measured lag, as a result of the large uncertainties on the lags at low RMS. Regardless, this low-RMS regime lies outside the relevant parameter space for our sample.

These results argue against a measurement bias being responsible for the observed correlation between RMS variability and lag excess, and instead point to a shared underlying cause; namely, the obscuring material traced by column density. This interpretation is supported by the multivariate regression, which showed that column density alone accounts for over 80\% of the variance in normalization excess. Once column density is included in the model, fractional RMS no longer contributes at a statistically significant level, suggesting that its apparent effect arises through covariance rather than a direct influence.

Although these findings strengthen the case for obscuration as the primary driver of the accretion disk size phenomenon, our analysis cannot establish causality. A fully causal interpretation would require assuming that column density is not itself confounded by some unmeasured variable that also influences the lag excess—a condition that, while plausible, cannot be verified with the current dataset. Nonetheless, our results are consistent with a scenario in which obscuring material directly shapes the observed lag behavior. This motivates the interpretation that follows, in which we explore how obscuration could give rise to the observed patterns across the sample and within individual sources. Future studies with larger samples and broader observational constraints will be necessary to rigorously assess causal relationships.

\subsection{Obscuration as the origin of the accretion disk size problem} \label{subsec:obscuration_discussion}

In recent years, the dominant interpretation for the accretion disk size problem was contamination from diffuse continuum emission in the broad-line region (BLR). This explanation was largely motivated by the shape of the lag excesses resembling the BLR emission, most prominently in the U-band ($\sim3500$\AA) near the Balmer jump. Photoionization modeling has shown that this emission is a natural consequence of BLR physics \citep{Korista_2001, Korista_2019, Netzer_2020, Netzer_2022}, and its presence is well-established in many systems. However, while BLR contamination is significant and likely a contributing factor, our results suggest it alone cannot account for the observed lag excesses. If it were, we would expect to see similar excesses in unobscured sources where the BLR is visible. Yet, the unobscured sources in our sample as a subpopulation show no significant deviation from the thin-disk predictions on average, even in the U band. However, it is worth mentioning that NGC~4593 is a clear exception to this: the HST lags demonstrate the excess around the Balmer jump that is not as significantly seen from the Swift data alone \citep{Cackett_2018}.

Another class of explanations has focused on developing new models to address shortcomings in how the standard thin disk model is implemented, particularly the assumptions and approximations involved in applying it to interpret observed lags. Recent works have proposed warped or structured disk geometries \citep{Starkey_2023}, as well as thermal reverberation models that incorporate general relativity, black hole spin, disk ionization, and the height of the corona \citep{Kammoun_2021a, Kammoun_2023, Langis_2024}. While these models can reproduce observed lag spectra, power spectra \citep{Panagiotou_2022a}, and SED \citep{Dovciak_2022}, they invoke physical processes intrinsic to the disk and its illumination and should therefore apply uniformly across the AGN population. One possible reconciliation is that both the increased lags and the obscuration arise from a common underlying driver, such as higher mass accretion rate ($\dot{M}$), which is expected to produce longer lags in these models via a more extended or puffed-up disk. However, we find no clear correlation between $\dot{M}$ and the observed lag excesses in our sample, and prior studies suggest the opposite trend--—higher accretion rates are typically associated with lower obscuration \citep[e.g.,][]{Ricci_2017}. This argues against $\dot{M}$ as a common driver of both phenomena in our case. The fact that the lag discrepancy appears only in the obscured subgroup argues against a universal structural shortcoming in the standard disk model and instead points to an extrinsic, line-of-sight dependent cause.

Our results instead indicate that obscuring material along the line of sight is likely reprocessing the incident radiation itself, producing additional delayed emission. The effective radius of reprocessing is shifted outwards to produce longer lags that are expected to scale with the degree of obscuration, as observed. The alternative that obscuration blocks emission from the inner disk, thereby shifting the effective reprocessing radius outward, is difficult to reconcile with the lack of significant obscuration of the UV/optical continuum.

This framework also helps contextualize prior observations. For example, the most extreme lag excess seen to-date--in Mrk~335, where the lag normalization exceed predictions by a factor of 12 \citep{Kara_2022}--can now be understood on a larger scale by this source having the highest measured column density of the campaign sources. Similarly, the dramatic evolution of lags in Mrk~817 during the AGN~STORM~2 campaign is naturally explained by changes in obscuration. At the start of the campaign, the source exhibited significant obscuration due to a multi-phase disk wind \citep{Kara_2021, Partington_2023, Zaidouni_2024}, and the measured UVOIR lags were much longer than expected. Midway through the campaign, however, the obscuration abated as the X-ray column density dropped by a factor of two. In this low-obscuration phase, the lag amplitudes halved, leading to a $>10\sigma$ decrease in the best-fit lag normalization. When the low column density window was isolated, the lags agreed with standard thin-disk predictions, including a resolution of the U-band excess \citep{Lewin_2024}. As the wind re-emerged late in the campaign, the column density returned to its earlier high level, and the lags lengthened again, closely matching those seen at the beginning of the campaign.

This repeating duty cycle demonstrates a clear link between column density and lag normalization within a single source that can be explained by our proposed interpretation: during high-obscuration phases, the obscuring material itself acts as a secondary reprocessor, introducing additional delayed emission as the photoionized gas emits both continuum and line radiation.

In any case, the conclusion that reprocessing in obscuring material explains the lag excess remains qualitative. More detailed theoretical modeling and simulations will be needed to quantify how much additional lag can be generated for different obscuration geometries and dust content. While the re-emission scenario may play a key role in the presence of photoionized disk wind, future work is warranted on whether the re-emission scenario holds up across the disk environments for the rest of the broader obscured AGN population. Mrk~817, in particular, offers a valuable case study for probing the potential causal connection between the disk structure, the obscuring material, and the resulting lags. While population-level studies are essential for identifying broad trends, targeted, high-cadence reverberation mapping of individual AGN in different states (e.g., with significantly varying line-of-sight column densities or accretion rates) can reveal how the lags evolve within a single source. Re-observing these systems in different epochs will be key to disentangling the relative roles of geometry, reprocessing, and obscuration in driving the lag excesses.

\subsection{X-ray/UV Timing Behavior and the Role of Obscuration}
\begin{figure*}[t!]
    \centering
    \includegraphics[width=\textwidth]{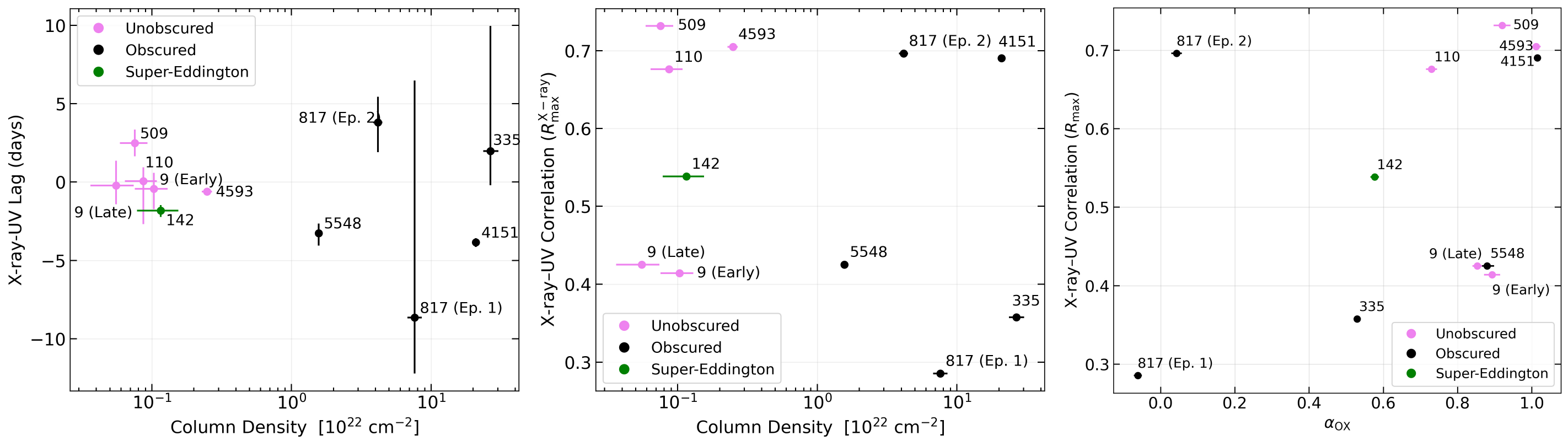}
    \caption{Relationships between X-ray/UV timing properties and physical parameters. \textit{Left:} X-ray to UV (UVW2, 1928\AA) lag amplitude versus column density ($N_{\rm H}$). The obscured subgroup exhibits notably larger scatter in lag amplitude compared to unobscured sources. \textit{Middle:} Maximum linear correlation between the X-ray and UVW2 light curves ($R_{\rm max}^{\rm X-ray}$) versus column density. \textit{Right:} $R_{\rm max}^{\rm X-ray}$ versus the optical–to–X-ray spectral slope ($\alpha_{\rm ox}$). These plots explore whether the coherence between the X-ray and UV emission---and the lag between them---depends on absorption or accretion properties.
}
    \label{fig:xray}
\end{figure*}

In addition to the accretion disk size problem in the UVOIR bands, the X-ray to UV lags and correlations give rise to another puzzle; namely, the inconsistent behavior of the X-ray to UV lags, with some sources showing the X-rays leading as expected, while others show the UV leading. This inconsistency, along with the often surprisingly low correlation between the X-ray and UV light curves, challenges the standard reprocessing paradigm in which high-energy X-ray photons from the corona are absorbed and re-emitted by the accretion disk at longer wavelengths.

In our sample, we find no statistically significant correlations between the maximum linear correlation coefficients of X-ray and UV bands ($R_{\rm max}^{\rm X-ray}$) and any of the measured physical parameters. Some relationships involving the X-ray to UV lags approached significance, with their confidence intervals overlapping the threshold/critical value, but large uncertainties in many of the lag measurements prevented us from drawing firm conclusions. In Figure~\ref{fig:xray}, we present the X-ray to UV (UVW2, 1928~\AA) lag amplitude versus column density, $R_{\rm max}^{\rm X-ray}$ versus column density, and $R_{\rm max}^{\rm X-ray}$ versus mass accretion rate.

The X-ray to UV lags in unobscured sources exhibit low scatter and are generally consistent with the small, negative lags predicted by thin-disk reprocessing (i.e., X-rays leading UV), with the exception of Mrk~509. For example, assuming a black hole mass of $M = 10^6M_\odot$ and an Eddington ratio of $\dot{m} = 0.05$, the expected X-ray–UVW2 lag is approximately $-0.01$ days, based on Equation~12 of \citet{Fausnaugh_2016}. In contrast, the obscured subgroup exhibits far greater scatter, with some sources showing large positive or negative lags and none showing consistent lags with simple corona-disk reprocessing expectations. This spread echoes previous surprising results, such as the (soft) X-rays delayed relative to the UV by over 12 days in Mrk~335, which was attributed to distant reflection that dominated the soft X-ray spectrum \citep{Parker_2019, Lewin_2023}. One possible explanation is that obscuration introduces additional complications in the X-ray/UV lag signal, such as reprocessing or scattering by outflows or winds. Indeed, modeling work such as \citet{Juranova_2022} has shown that variable photoionized outflows can produce complex lag behavior and reduce coherence through non-linear ionization responses. We might expect to see this behavior from, for example, the photoionized disk winds seen in Mrk~817 and NGC~5548, but additional work is needed for modeling the feasibility of this within the obscured subgroup.

A similar pattern emerges in the degree of correlation between the X-ray and UV. With the exception of Fairall~9, all of the unobscured sources show high X-ray/UV correlation, as expected from a relatively clear line of sight to the disk. It is interesting that Fairall~9 shows the largest change in the lag normalization across its campaign, transitioning from negative to positive values and setting the range measured in our sample ($\tau_0 = -0.6$ to $1.24\text{ days}$). Unlike the changing lags seen in Mrk~817, where such changes correlated to variable obscuration, Fairall~9 is unobscured, suggesting these changes might instead result from variations in the corona and/or accretion flow themselves. However, \citet{Hagen_2024} suggest that the lags in Fairall~9 are indeed due to emission from a disk wind that is not along our line of sight. In general, this interpretation could account for the low X-ray/UV correlation, as \citet{Panagiotou_2022b} showed that dynamic variability in the X-ray source can naturally suppress coherence between the bands.

Similar to the lags, the obscured subpopulation shows significant variation in X-ray/UV correlation. In Mrk~817, the X-ray/UV correlation was low during times of high column density in the campaign (Epoch~1, $R < 0.3$), and was high during the less obscured Epoch~2 ($R \sim 0.7$). In this case, the obscuration appeared to suppress correlation via blocking the direct X-ray flux. On the other hand, NGC~4151, one of the most heavily obscured sources in our sample, shows a surprisingly high X-ray/UV correlation. Here, the X-ray flux remains visible, but does not appear to track the UV variability in a simple way. This behavior reflects more complicated dynamics, such as changes in the ionization state of the obscuring material \citep{Juranova_2022}, or that the reprocessing environment is complex even within the outer disk, leading to multiple layers of reprocessing that might break the simple causal connection between X-rays and UV. For instance, \citet{Miller_2018, Zoghbi_2019, xrism_2024}, found that the narrow Fe~K$\alpha$ line in NGC~4151 likely arises from intermediate radii in the BLR or a warped disk, with evidence for structural changes between high and low flux states that suggest structural changes in the reprocessing region.

One additional possibility is that differences in $R_{\rm max}$ among obscured sources may stem from variable obscuration. If the obscurer is dust-free, it could heavily modulate the X-ray flux while leaving the UV relatively unaffected, lowering the correlation. However, if the obscuring material contains significant dust, it could also suppress the UV flux via extinction, maintaining a high $R_{\rm max}$ even during high $N_{\rm H}$ intervals. This may be relevant in the case of the dust in NGC~4151 \citep[e.g.,][]{Lyu_2021}. This diversity within the obscured subgroup suggests that the dynamical evolution of the surrounding material may  play a critical role in shaping the observed timing properties, even between AGN with comparable levels of obscuration.

Lastly, we find no compelling trend between $R_{\rm max}^{\rm X-ray}$ and accretion rate. Mrk~142, the only super-Eddington source in our sample, shows a only moderate correlation, whereas the rest of the sources at similarly low accretion rates span a wide range of correlation values. This runs counter to the idea that low X-ray to UV correlations is the byproduct of low X-ray luminosity and, by extension, low Eddington ratios. If this were the case, we would expect lower accretion rates to correspond to systematically lower coherence, which we do not observe.

\section{Conclusion}
We analyzed the X-ray spectra and UVOIR interband lags for the nine AGN targeted by recent high-cadence, multi-wavelength reverberation mapping campaigns. The main results are as follows:

\begin{itemize}
\item We modeled the Swift X-ray spectra, which show two main behaviors: smooth, power-law-like continua consistent with unobscured coronal emission, and heavily curved, absorbed spectra indicative of significant line-of-sight obscuration.

\item Using a Gaussian Mixture Model based on X-ray column density and hardness ratio, we classified the sources into two statistically distinct groups with high confidence in their assignments ($>99.7\%$), consistent with visual inspection.

\item The measured lag normalizations are on-average $3.45 \pm 0.27$ times larger than predicted by standard thin-disk theory, showing the significance of the well-known accretion disk size problem.

\item Our results show that the accretion disk size problem is not a universal feature of AGN, but instead arises in a subpopulation of systems with substantial obscuration. Based on subgroup averages, the observed lag amplitudes exceeded thin-disk predictions only among the obscured AGN: unobscured sources had a mean normalization of $1.00 \pm 0.31$, consistent with theoretical expectations, while obscured AGN showed significantly larger mean values ($5.21 \pm 0.47$, $p = 0.008$).

\item Both X-ray column density and UVW2-band RMS variability showed strong correlations with the lag normalization excess. However, multivariate regression revealed that column density alone explains over 80\% of the variance in excess lags and remains the only statistically significant predictor when accounting for shared variance, indicating that the apparent link with RMS variability is likely a byproduct of its covariance with column density.

\item These results support the interpretation that obscuring material is the key driver of the accretion disk size phenomenon by introducing additional reprocessed emission from the absorbing material itself. It is difficult to reconcile the alternative idea that obscuration is blocking emission from the inner disk, thus shifting the effective reprocessing radius outward, given that we do not see significant obscuration of the UV/optical continuum.

\item The X-ray to UV timing behavior reinforces this picture: all but one unobscured AGN show small, negative X-ray-to-UV lags, while those in the obscured AGN show much larger scatter. This diversity likely reflects a combination of factors in these sources, including suppression of the X-ray flux, complicated structural or ionization changes of the inner reprocessing regions.

We emphasize that our sample is relatively small (nine AGN, with some repeat epochs). While the trends are robust within this dataset, confirmation with larger, more diverse samples will be important. In addition, for some sources such as Fairall 9 and Mrk 817, the measured lags depend strongly on the campaign epoch, highlighting that temporal variability within individual AGN may influence trends across the population at large. These results imply that lag-based accretion disk sizes in unobscured AGN strongly support thin-disk theory, while past measurements in mixed samples may have been systematically biased high. Future reverberation mapping campaigns should therefore explicitly account for obscuration in both their design and interpretation.
\end{itemize}

\begin{acknowledgments}
CL and EK acknowledged NASA grant 80NSSC22K1120 for support. CP acknowledges NASA grant 80NSSC25K7404 for support. This work made us of observations obtained with the Hubble Space Telescope, which is operated by the Space Telescope Science Institute, under the auspices of the Association of Universities for Research in Astronomy, Inc., under NASA contract NAS5-26555.
\end{acknowledgments}

\facilities{HST, Swift, LCOGT, Liverpool:2m, Wise Observatory, Zowada, CAO:2.2m, YAO:2.4m}
\software{STELA Toolkit \citep{}}

\appendix
\begin{deluxetable*}{lccccc}
\tablecaption{Best-fit spectral parameters for each AGN with 90\% confidence intervals.}
\label{table:spectral_params}
\tablehead{
\colhead{Source} &
\colhead{$N_{\mathrm{H}}$ ($10^{22}$ cm$^{-2}$)} &
\colhead{Covering Fraction} &
\colhead{Photon Index} &
\colhead{$kT$ (eV)} &
\colhead{$\chi^2 / \rm dof$}
}
\startdata
Mrk 110           & $0.09^{+0.04}_{-0.04}$  & --                        & $1.71^{+0.03}_{-0.03}$ & $103.7^{+4.9}_{-4.0}$  & $763.33 / 641$ \\
Mrk 142           & $0.12^{+0.06}_{-0.06}$  & --                        & $2.25^{+0.07}_{-0.07}$ & $94.9^{+7.5}_{-6.0}$   & $275.48 / 283$ \\
Mrk 509           & $0.08^{+0.03}_{-0.03}$  & --                        & $1.71^{+0.02}_{-0.02}$ & $89.8^{+3.2}_{-2.8}$   & $651.17 / 638$ \\
NGC 4593          & $0.25^{+0.03}_{-0.03}$  & --                        & $1.74^{+0.03}_{-0.03}$ & $79.9^{+1.9}_{-1.8}$   & $714.22 / 613$ \\
Fairall 9 (Early) & $0.10^{+0.04}_{-0.05}$  & --                        & $1.90^{+0.03}_{-0.03}$ & $100.8^{+5.9}_{-4.5}$  & $538.04 / 503$ \\
Fairall 9 (Late)  & $0.06^{+0.10}_{-0.03}$  & --                        & $1.92^{+0.03}_{-0.03}$ & $107.0^{+6.0}_{-4.5}$  & $575.75 / 551$ \\
Mrk 335           & $27.88^{+2.27}_{-2.04}$ & $0.84^{+0.01}_{-0.01}$    & $1.69^{+0.03}_{-0.03}$ & $149.8^{+3.6}_{-3.6}$  & $246.16 / 192$ \\
Mrk 817 (Epoch 1) & $7.87^{+0.68}_{-0.60}$  & $0.79^{+0.01}_{-0.01}$    & $1.42^{+0.02}_{-0.02}$ & $117.0^{+7.0}_{-6.9}$  & $173.56 / 190$ \\
Mrk 817 (Epoch 2) & $4.20^{+0.23}_{-0.21}$  & $0.76^{+0.01}_{-0.01}$    & $1.62^{+0.02}_{-0.02}$ & $87.5^{+2.5}_{-2.5}$   & $335.28 / 355$ \\
NGC 5548          & $1.56^{+0.02}_{-0.02}$  & $0.93^{+0.01}_{-0.01}$    & $1.46^{+0.01}_{-0.01}$ & $89.9^{+1.2}_{-1.2}$   & $775.22 / 668$ \\
NGC 4151          & $21.10^{+0.22}_{-0.21}$ & $0.95^{+0.01}_{-0.01}$    & $1.20^{+0.01}_{-0.01}$ & $100.6^{+1.6}_{-1.6}$  & $1382.78 / 828$ \\
\enddata
\end{deluxetable*}

\section{Computing Eddington Ratios from X-ray Luminosity} \label{sec:appendix_bolometric}

\begin{figure*}[t!]
    \centering
    \includegraphics[width=0.72\textwidth]{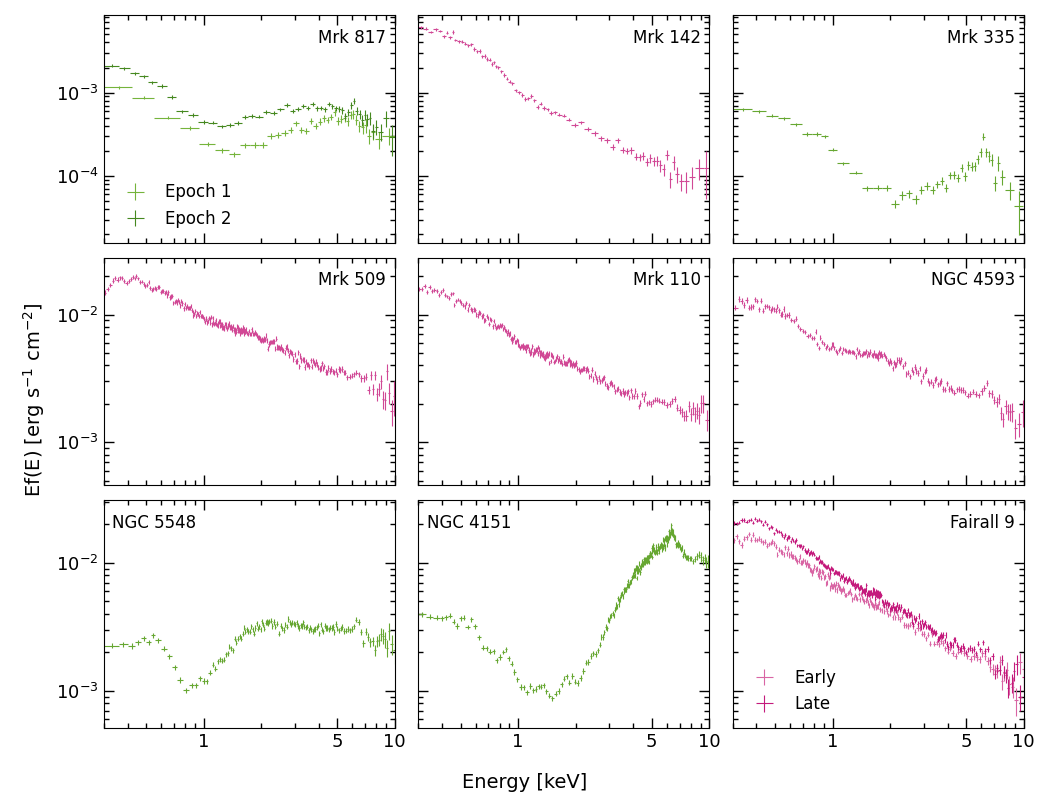}
    \caption{
    Per–source \textit{Swift} X-ray spectra (0.3–10\,keV), with unobscured spectra in pink and obscured in green. For Mrk~817 and Fairall~9, two epochs are over-plotted. Pink spectra exhibit smooth, power-law--like continua, whereas green panels show pronounced curvature (particularly $\sim$1--4\,keV) characteristic of absorption along the line of sight.
    }
    \label{fig:spectra_grid}
\end{figure*}

In the main text, we adopt Eddington ratios from the literature because, in many cases, these studies do modeling of the spectral energy distribution (SED) instead of scaling. However, this does mean that the accretion rates are calculated in a heterogeneous way, which may cause biases in our correlation study. To test how robust our results are given this choice, we repeat our analysis using a consistent bolometric correction method instead. Specifically, we compute the bolometric luminosity by applying a luminosity-dependent bolometric correction to the intrinsic (absorption-corrected) 2--10 keV X-ray luminosity 2--10 keV X-ray luminosity, with fluxes obtained via the fits using \texttt{cflux}. This correction factor, $K_X$, depends on the X-ray luminosity \citep[computed using Equation 3 in][]{Duras_2020}. The resulting bolometric luminosity is then used to compute the Eddington ratio $L_{\mathrm{bol}} / L_{\mathrm{Edd}}$. 

This procedure results in slightly different expected lags from the standard disk model (because the expected lag depends on accretion rate), and thus slightly different lag excesses. The values used for this test and their comparison to the original literature-based Eddington ratios are listed in Table~\ref{tab:eddington_comparison}. This method yields a notably lower Eddington ratio for the value of Mrk~142 from \citet{Cackett_2020} using the optical luminosity at 5100~\AA \citep[using equation 2 in][]{Du_2015}, although they similarly find a wide range of values depending on method. Using the bolometric luminosity results in an Eddington ratio closer to 0.6, which is comparable to that of Mrk~110. This value no longer supports placing Mrk~142 in a separate super-Eddington class, however it is important to note that this bolometric correction is likely not appropriate for super-Eddington systems \citep[see, e.g. the recent work of][]{Pacucci_2025}. For the purposes of this analysis, we therefore classify Mrk~142 as unobscured, in agreement with its classification by the GMM.

To evaluate the significance of the lag excesses in each subgroup, we use $t$-tests that we justify by checking for approximate normality. The Shapiro–Wilk test returns a test statistic of 0.90 and a $p$-value of 0.17, while the Anderson–Darling test yields a statistic of 0.49, below the 10\% critical threshold of 0.57. As a result, both tests indicate that the assumption of normality is not strongly violated.

For the full sample, we find a mean lag excess of $4.35 \pm 1.34$, yielding a $p$-value of 0.016 from a one-sided $t$-test (testing $H_0$: mean excess $= 1$ vs. $H_1$: mean excess $> 1$). Among the obscured sources, the mean is $6.27 \pm 1.53$ with $p = 0.013$, indicating strong evidence for the excess lags. The unobscured sources yield a mean of $1.9 \pm 1.2$ and a not-significant $p$-value of 0.23, again showing consistency with expectations from the standard accretion disk model. These results are consistent with the conclusions in the main text: the full and obscured samples show statistically significant deviations from the expected lag, while the unobscured sources remain consistent with thin-disk predictions.

\begin{table}[ht]
\centering
\caption{Comparison of Eddington ratios computed using bolometric corrections (\textit{this work}) versus literature values. The bolometric luminosity is computed using a luminosity-dependent correction $K_X$ applied to the intrinsic 2--10~keV luminosity, following \citet{Duras_2020}.}
\label{tab:eddington_comparison}
\begin{tabular}{lccc}
\hline
\hline
Source & $K_X$ & $\dot{m_E}$ (This Work) & $\dot{m_E}$ (Literature) \\
\hline
Mrk~335                  & 15.6  & 0.025 & 0.070 \\
Mrk~817 (Epoch 1)  & 16.1  & 0.047 & 0.200 \\
Mrk~817 (Epoch 2)   & 16.3  & 0.061 & 0.200 \\
NGC~5548                & 16.5  & 0.067 & 0.050 \\
Mrk~110                 & 17.9  & 0.539  & 0.650 \\
NGC~4151                & 15.7  & 0.028 & 0.010 \\
Mrk~142                 & 15.9  & 0.573  & 3.400 \\
Mrk~509                 & 18.9  & 0.160  & 0.050 \\
NGC~4593                & 15.7  & 0.083 & 0.080 \\
Fairall~9 (Early)              & 18.9  & 0.070 & 0.035 \\
Fairall~9 (Late)        & 19.3  & 0.082 & 0.035 \\
\hline
\end{tabular}
\end{table}

\bibliography{bibliography}{}
\bibliographystyle{aasjournalv7}
\end{document}